\documentclass{ws-rmp}

\usepackage{graphicx}
\usepackage{amsfonts}
\usepackage{curves}
\usepackage{amstext}
\usepackage{bm}
\usepackage{rotating}
\usepackage{multirow}
\usepackage{color}
\usepackage{hyperref}
\usepackage[square,comma,numbers,sort&compress]{natbib}

\newcommand{\fl}{}  

\def\x{\bm{x}}

\def\PP{\mathcal P}
\def\E{{\mathbb E}}
\def\P{{\mathbb P}}
\def\R{{\mathbb R}}
\def\Z{{\mathbb Z}}

\def\I{{\mathcal I}}

\def\K{{\mathcal K}}
\def\L{{\mathcal L}}

\def\W{{\mathcal W}}

\def\pa{\partial\Omega}

\def\ctan{\mathrm{ctan}}

\def\ve{\varepsilon}

\newcommand{\clr}{\color{red}}

\def\C{{\mathbb C}}

\def\LLambda{\ell}

\begin{document}

\title{A physicist's guide to explicit summation formulas involving zeros of Bessel functions and related spectral sums}
\author{Denis S. Grebenkov}
\address{Laboratoire de Physique de la Mati\`ere Condens\'ee, \\
CNRS -- Ecole Polytechnique, IP Paris, F-91128 Palaiseau, France 
 \email{denis.grebenkov@polytechnique.edu}}

\maketitle

\begin{history}
\received{(\today)}
\end{history}

\begin{abstract}
In this pedagogical review, we summarize the mathematical basis and
practical hints for the explicit analytical computation of spectral
sums that involve the eigenvalues of the Laplace operator in simple
domains such as $d$-dimensional balls (with $d = 1,2,3$), an annulus,
a spherical shell, right circular cylinders, rectangles and
rectangular cuboids.  Such spectral sums appear as spectral expansions
of heat kernels, survival probabilities, first-passage time densities,
and reaction rates in many diffusion-oriented applications.  As the
eigenvalues are determined by zeros of an appropriate linear
combination of a Bessel function and its derivative, there are
powerful analytical tools for computing such spectral sums.  We
discuss three main strategies: representations of meromorphic
functions as sums of partial fractions, Fourier-Bessel and Dini
series, and direct evaluation of the Laplace-transformed heat kernels.
The major emphasis is put on a pedagogic introduction, the practical
aspects of these strategies, their advantages and limitations.  The
review gathers many summation formulas for spectral sums that are
dispersed in the literature.
\end{abstract}

\noindent{\it Keywords\/}: Laplacian eigenvalues, Bessel functions, spectral sums, summation formulas, heat kernel, diffusion



\section{Introduction}
\label{sec:intro}

The Laplace operator, $\Delta = \nabla^2$, plays the central role in
mathematical physics and found countless applications in quantum
mechanics, acoustics, chemical physics, biology, heat transfer,
hydrodynamics, nuclear magnetic resonance, and stochastic processes
\cite{Carslaw,Crank,Redner,Bass,Borodin,Cole,Grebenkov07,Grebenkov09,Bressloff13}.
In a bounded Euclidean domain $\Omega \subset \R^d$, the Laplace
operator has a discrete spectrum, i.e., a countable number of
eigenvalues $\lambda_m$ and eigenfunctions $U_m$ satisfying
\begin{eqnarray}
\Delta U_m + \lambda_m U_m &=& 0  \quad \mbox{in}~\Omega, \\
\label{eq:BC}
\LLambda \frac{\partial}{\partial n} U_m + U_m &=& 0  \quad \mbox{on}~ \pa , 
\end{eqnarray}
subject to Dirichlet ($\LLambda = 0$), Neumann ($\LLambda = \infty$)
or Robin ($0 < \LLambda < \infty$) boundary condition on a smooth
boundary $\pa$, with $\partial/\partial n$ being the normal derivative
oriented outward the domain $\Omega$ (the Neumann case can formally be
understood as the limit $\LLambda \to \infty$; in particular, the
boundary condition (\ref{eq:BC}) reads $\partial U_m/\partial n = 0$).
The eigenvalues are nonnegative and grow to infinity, $0 \leq
\lambda_1 \leq \lambda_2 \leq \ldots \nearrow +\infty$, whereas the
eigenfunctions form an orthonormal complete basis in the space
$L_2(\Omega)$ of square integrable functions.  In particular, the
eigenvalues and eigenfunctions determine the Green's functions of
homogeneous diffusion and wave equations and thus allow for solving
most common boundary value problems \cite{Gardiner,Courant,Port}.  For
instance, the spectral decomposition
\begin{equation}   \label{eq:prop_Laplace}
P(\x,t|\x_0) = \sum\limits_{m=1}^\infty U_m(\x) \, U_m^*(\x_0) \, e^{-Dt\lambda_m}
\end{equation}
determines the Green's function of the diffusion (or heat) equation,
\begin{equation}
\frac{\partial}{\partial t} P(\x,t|\x_0) = D \Delta P(\x,t|\x_0)  \quad \textrm{in} ~\Omega,
\end{equation}
subject to the same boundary condition as Eq. (\ref{eq:BC}) and the
initial condition $P(\x,t=0|\x_0) = \delta(\x-\x_0)$, where $\delta$
is the Dirac distribution, $D$ is the diffusion coefficient, $\x_0$ is
a fixed point in $\Omega$, and asterisk denotes the complex conjugate.
This Green's function, also known as heat kernel or propagator,
describes various properties of diffusion-controlled processes in
reactive media \cite{Rice,Torquato91,Galanti16,Grebenkov19}.  In turn,
the Laplace transform of the heat kernel,
\begin{equation}  \label{eq:prop_Laplace2}
\tilde{P}(\x,p|\x_0) = \int\limits_0^\infty dt \, e^{-pt} \, P(\x,t|\x_0) = 
\sum\limits_{m=1}^\infty \frac{U_m(\x) \, U_m^*(\x_0)}{p + D\lambda_m} \,,
\end{equation}
is the Green's function of the modified Helmholtz equation,
\begin{equation}  \label{eq:Helm}
\bigl(p - D \Delta\bigr) \tilde{P}(\x,p|\x_0) = \delta(\x - \x_0)  \quad \textrm{in} ~\Omega,
\end{equation}
subject to the same boundary condition as Eq. (\ref{eq:BC}), and tilde
denotes the Laplace transform.

The spectral decomposition (\ref{eq:prop_Laplace}) and formulas
derived from it determine most diffusion-based characteristics.  
An emblematic example is the heat trace,
\begin{equation}  \label{eq:spectral_HT_exp}
Z(t) = \int\limits_\Omega d\x_0 \, P(\x_0,t|\x_0) = \sum\limits_{m=1}^\infty e^{-D \lambda_m t} ,
\end{equation}
which can be interpreted as the probability density of the return to
the starting point $\x_0$, averaged over the starting point.  This
quantity is also known as the return-to-the-origin probability
\cite{Mitra95,Schwartz97,Grebenkov09b}.  Its Laplace transform,
\begin{equation}  \label{eq:spectral_HT}
\tilde{Z}(p) = \int\limits_0^\infty dt \, e^{-pt} \, Z(t) = \sum\limits_{m=1}^\infty \frac{1}{p + D\lambda_m} \,,
\end{equation}
is one of the simplest spectral sums.

Another important example is the survival probability of a particle
diffusing in a domain with partially absorbing boundaries:
\begin{equation}  \label{eq:St}
\fl
S(t|\x_0) = \P_{\x_0}\{ \tau > t\} = \int\limits_\Omega d\x \, P(\x,t|\x_0) 
= \sum\limits_{m=1}^\infty e^{-Dt\lambda_m} \, U_m^*(\x_0)  \int\limits_\Omega d\x \, U_m(\x),
\end{equation}
from which the probability density of the (random) reaction time
$\tau$ (i.e., the first-passage time to a reaction event) follows as
$H(t|\x_0) = -\partial S(t|\x_0)/\partial t$.  If the starting point
is uniformly distributed (or, equivalently, the initial concentration
of particles is uniform), the volume average of the survival
probability yields the NMR signal attenuation due to surface
relaxation \cite{Brownstein79}:
\begin{equation}  \label{eq:St_av}
\fl
\overline{S(t)} = \frac{1}{|\Omega|}\int\limits_{\Omega} d\x_0 \, S(t|\x_0) =
\frac{1}{|\Omega|} \sum\limits_{m=1}^\infty e^{-Dt\lambda_m} \left|\int\limits_\Omega d\x \, U_m(\x) \right|^2 ,
\end{equation}
whereas the volume average of the probability density $H(t|\x_0)$
gives the reaction rate on the boundary $\pa$ of the domain
\cite{Grebenkov19f}:
\begin{equation}  \label{eq:Ht_av}
\fl
\overline{H(t)} = \frac{1}{|\Omega|} \int\limits_{\Omega} d\x_0 \, H(t|\x_0) =
\frac{D}{|\Omega|} \sum\limits_{m=1}^\infty e^{-Dt\lambda_m} \, \lambda_m \, \left|\int\limits_\Omega d\x \, U_m(\x) \right|^2 .
\end{equation}
The Laplace transform of these spectral expansions yields several
spectral sums that are often used to investigate the short-time
asymptotic behavior of the above quantities.  For instance, one gets
from (\ref{eq:St})
\begin{equation}  \label{eq:Sp}
\fl
\tilde{S}(p|\x_0) = \sum\limits_{m=1}^\infty \frac{1}{p + D\lambda_m} \, U_m^*(\x_0)  \int\limits_\Omega d\x \, U_m(\x).
\end{equation}
In particular, the mean reaction time reads
\begin{equation}  
\E_{\x_0}\{ \tau \} = \tilde{S}(0|\x_0) = \frac{1}{D} \sum\limits_{m=1}^\infty \lambda_m^{-1} \, U_m^*(\x_0) \int\limits_\Omega d\x \, U_m(\x),
\end{equation}
while its volume average is
\begin{equation}
\overline{\E\{ \tau \}} = \frac{1}{D\, |\Omega|} \sum\limits_{m=1}^{\infty} \lambda_m^{-1} \, \biggl| \int\limits_\Omega d\x \, U_m(\x) \biggr|^2.
\end{equation}

Yet another common quantity of interest in the effective diffusion
coefficient in a confining domain:
\begin{eqnarray}  \nonumber
D(t) &=& \frac{1}{2d} \int\limits_\Omega d\x_0 \int\limits_\Omega d\x \, |\x - \x_0|^2 \, P(\x,t|\x_0) \\  
&=& \frac{1}{2d} \sum\limits_{m=1}^{\infty} e^{-Dt\lambda_m} 
\int\limits_\Omega d\x_0 \int\limits_\Omega d\x \, |\x - \x_0|^2 \, U_m^*(\x_0) \, U_m(\x).
\end{eqnarray}
More generally, the $j$-point correlation function of the particle
position $X_t$ is
\begin{align*}  \nonumber
& C(t_1,\ldots,t_j|\x_0) = \E_{\x_0}\{X_{t_1} \, X_{t_2} \ldots  X_{t_j}\} = \int\limits_\Omega d\x_1 \ldots \int\limits_\Omega d\x_j \,
\x_j   \\  
& \quad \times P(\x_j,t_j-t_{j-1}|\x_{j-1})\, \x_{j-1} \, P(\x_{j-1},t_{j-1}-t_{j-2}|\x_{j-2}) \ldots \x_1 \, P(\x_1,t_1|\x_0)
\end{align*}
(with $0 < t_1 < \ldots < t_j$), which can be expressed as the
$j$-fold spectral sum:
\begin{align} \nonumber
& C(t_1,\ldots,t_j|\x_0) 
= \sum\limits_{m_1=1}^\infty \ldots \sum\limits_{m_j=1}^\infty 
e^{-D\lambda_{m_1} t_1 - D\lambda_{m_2} (t_2-t_1) - \ldots - D\lambda_{m_j} (t_j - t_{j-1})} \\    \label{eq:Xnt}
& \qquad \times  U_{m_1}^*(\x_0) \, B_{m_1,m_2} \, B_{m_2,m_3}  \ldots B_{m_{j-1},m_j} \, \int\limits_{\Omega} d\x_j \, \x_j \, U_{m_j}(\x_j) , 
\end{align}
where
\begin{equation*}
B_{m,m'} = \int\limits_{\Omega} d\x \, U_m(\x) \, \x \, U_{m'}^*(\x).
\end{equation*}
To evaluate this multiple sum, one needs to perform $j$ Laplace
transforms, for each time variable $t_i$:
\begin{align} \nonumber
& \tilde{C}(p_1,\ldots,p_j|\x_0) = \sum\limits_{m_1=1}^\infty \ldots \sum\limits_{m_j=1}^\infty 
\frac{1}{p_1 + D(\lambda_{m_1} - \lambda_{m_2})} \, \ldots 
\, \frac{1}{p_j + D\lambda_{m_j}} \\    \label{eq:Xnp}
& \qquad \times  U_{m_1}^*(\x_0) \, B_{m_1,m_2} \, B_{m_2,m_3}  \ldots B_{m_{j-1},m_j} \, \int\limits_{\Omega} d\x_j \, \x_j \, U_{m_j}(\x_j) .
\end{align}
Similar expressions appear in the perturbative expansion of 
the diffusion NMR signal attenuation in a linear magnetic field
gradient \cite{Grebenkov07,Mitra93,Grebenkov16}.  As the eigenvalues
and eigenfunctions are in general not known, only the asymptotic
behavior ofspectral sums is usually accessible, e.g., from the Weyl's
asymptotic law for eigenvalues or from the analysis of the heat kernel
\cite{Davies,Gilkey,Grebenkov13}.

\begin{figure}
\begin{center}
\includegraphics[width=100mm]{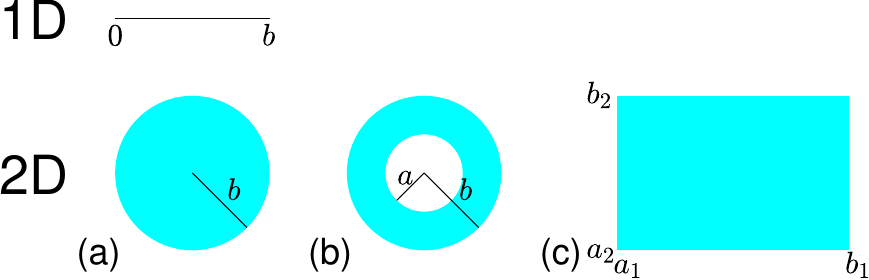} 
\includegraphics[width=102mm]{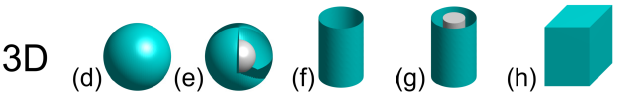} 
\end{center}
\caption{
Simple domains for which the Laplacian eigenvalues and eigenfunctions
involve the zeros of sine, Bessel and spherical Bessel functions
and/or their derivatives.  For an interval, a disk (a), a circular
annulus (b), a ball (d) and a spherical shell (e), many spectral sums
can be computed exactly by summation techniques discussed in the
review.  For a rectangle (c), a right circular cylinder (f), a
cylindrical shell (g), and a rectangular cuboid (h), each eigenvalue
is the sum of eigenvalues in former basic domains, e.g., $\lambda_{nk}
= \pi^2 n^2/(b_1-a_1)^2 + \pi^2 k^2/(b_2-a_2)^2$ ($k,n=1,2,\ldots$)
for the rectangle with Dirichlet boundary conditions; as a
consequence, spectral sums over double (or triple) indices can be
simplified by summation techniques.}
\label{fig:domains}
\end{figure}

However, for a very limited number of simple domains (see examples in
Fig. \ref{fig:domains}), their symmetries allow for the separation of
variables, and the eigenvalues and eigenfunctions are known explicitly
\cite{Carslaw,Crank}.  An interval, a disk, a circular annulus, a ball
and a spherical shell are the most studied examples
\cite{Thambynayagam} that play the role of important toy models for
understanding diffusion-controlled processes, e.g., in chemical
physics and biology.  The Laplacian eigenfunctions in these basic
domains, defined generally as $\Omega = \{ \x \in \R^d ~:~ a < |\x| <
b\}$ with given radii $0 \leq a < b$, are
\begin{equation}  \label{eq:Um}
\fl
\begin{array}{r l l}
U_k(r) & = A_k \, \cos(\sqrt{\lambda_k} r) + B_k \, \sin(\sqrt{\lambda_k} r)  & (d = 1),  \\
U_{nk}(r,\phi) & = \bigl(A_{nk} \, J_n(\sqrt{\lambda_{nk}} r) + B_{nk} \, Y_n(\sqrt{\lambda_{nk}} r) \bigr) \, e^{in\phi} & (d = 2), \\
U_{nkl}(r,\theta,\phi) & = \bigl(A_{nk} \, j_n(\sqrt{\lambda_{nk}} r) + B_{nk} \, y_n(\sqrt{\lambda_{nk}} r)\bigr)\, 
P_n^l(\cos\theta) \, e^{il\phi} & (d = 3), \\
\end{array}
\end{equation}
where $P_n^l(z)$ are the associated Legendre polynomials, $J_n(z)$ and
$Y_n(z)$ are the Bessel functions of the first and second kind, and
$j_n(z)$ and $y_n(z)$ are the spherical Bessel functions of the first
and second kind.  In two and three dimensions, polar and spherical
coordinates are respectively used, while double ($n$ and $k$) and
triple ($n$, $k$, and $l$) indices are employed to enumerate the
eigenfunctions and eigenvalues.  As this enumeration is formally
equivalent to that by the single index $m$, we will use both schemes
and choose one depending on the context.  For each eigenfunction $U_k$
(or $U_{nk}$, or $U_{nkl}$), the three unknown parameters, the
eigenvalue $\lambda_k$ (or $\lambda_{nk}$) and the coefficients $A_k$
and $B_k$ (or $A_{nk}$ and $B_{nk}$), are determined via one $L_2$
normalization condition,
\begin{equation}
\int\limits_\Omega d\x \, |U_m(\x)|^2 = 1 ,
\end{equation}
and two boundary conditions in Eq. (\ref{eq:BC}), written explicitly
as
\begin{equation}  \label{eq:Robin}
\bigl(\LLambda_a \partial_r U_m - U_m\bigr)_{r = a} = 
\bigl(\LLambda_b \partial_r U_m + U_m\bigr)_{r = b} = 0 ,
\end{equation}
with two given constants $\LLambda_a$ and $\LLambda_b$ (note the sign
change in the first relation due to opposite orientation of the radial
and normal derivatives).  The explicit form of the coefficients
$A_{nk}$ and $B_{nk}$, as well as an explicit equation determining the
eigenvalues $\lambda_k$ (or $\lambda_{nk}$) are well known and will be
summarized below.  Note that the radial functions in Eq. (\ref{eq:Um})
satisfy the radial eigenvalue problem $\L_d u + \lambda u = 0$ for the
differential operators
\begin{equation}  \label{eq:L_radial}
\L_1 = \partial_r^2 ,  \qquad
\L_2 = \partial_r^2 + \frac{1}{r} \partial_r - \frac{n^2}{r^2} \,, \qquad
\L_3 = \partial_r^2 + \frac{2}{r} \partial_r - \frac{n(n+1)}{r^2} \,.
\end{equation}

In this review, we discuss three strategies to compute {\it
explicitly} spectral sums involving the eigenvalues $\lambda_k$ (or
$\lambda_{nk}$): (i) representation of meromorphic functions as sums
of partial fractions (Sec. \ref{sec:meromorphic}), (ii) Fourier-Bessel
and Dini series (Sec. \ref{sec:Dini}), and (iii) direct evaluation of
Laplace-transformed heat kernels (Sec. \ref{sec:kernel}).  The first
strategy relies on the corollary of the classical Mittag-Leffler
theorem from complex analysis that affirms the existence of a
meromorphic function with prescribed poles and allows one to express
any meromorphic function as a sum of partial fractions.  In this way,
one aims at identifying an appropriate meromorphic function to
represent a given spectral sum.  The second strategy is based on the
theory of Fourier-Bessel and Dini series that generalize Fourier
series to Bessel functions.  Here, one searches for an appropriate
function to be decomposed as the Dini series whose form would coincide
with the studied spectral sum.  The third strategy employs the
spectral decomposition (\ref{eq:prop_Laplace2}) of the
Laplace-transformed heat kernel (or propagator).  As the left-hand
side can be evaluated independently by solving the modified Helmholtz
equation (\ref{eq:Helm}) in simple domains, one obtains a powerful
identity for evaluating spectral sums.  Remarkably, even though the
eigenvalues usually not known explicitly and require a numerical
solution of a transcendental equation, spectral sums like
Eq. (\ref{eq:prop_Laplace2}) involving such unknown eigenvalues can be
often computed exactly and explicitly.  Moreover, the particular form
(\ref{eq:Um}) of the eigenfunctions often enables analytical
computation of the integrals in (\ref{eq:St} -- \ref{eq:Xnp}) (or
similar) and thus allows one to derive explicit formulas for such
spectral sums.

The ability of computing spectral sums is particularly relevant
for understanding the short-time limit of diffusion-controlled
quantities.  In fact, as each eigenmode in the spectral expansion
(\ref{eq:prop_Laplace}) (or similar) is weighted by the exponential
factor $e^{-Dt\lambda_m}$, more and more terms are needed to determine
the behavior of the heat kernel and other quantities in this limit.
As a consequence, the analysis is usually performed in Laplace domain
by evaluating the spectral sum (\ref{eq:prop_Laplace2}), determining
its asymptotic behavior as $p\to \infty$, and applying Tauberian
theorems to get back to the time domain.

At the same time, spectral sums should not be exclusively associated
with the short-time limit.  For instance, the integral of the heat
trace in Eq. (\ref{eq:spectral_HT_exp}) yields the sum of inverse
eigenvalues, which also corresponds to the limit $p\to 0$ of
Eq. (\ref{eq:spectral_HT}).  In other words, such spectral sums
correspond to the long-time limit.  For instance, Neuman used the
summation formula (\ref{eq:eta_useful4}) involving the zeros of the
derivative of Bessel and spherical Bessel functions to derive the
long-time asymptotic behavior of the diffusion NMR signal
\cite{Neuman74}.  This analysis was further extended in
\cite{Grebenkov07} (see also more detailed discussions in
\cite{Grebenkov09,Grebenkov07b}).

The explicit summation formulas also help simplifying some analytical
solutions.  For instance, the separation of variables for the
diffusion equation in a right circular cylinder, $\Omega = \{ \x \in
\R^2 ~:~ |\x| < R\} \times (a,b) \subset \R^3$, would naturally lead
to double sums over the eigenvalues for an interval $(a,b)$ and the
eigenvalues for a disk of radius $R$ (this is also true for a hollow
cylinder when the disk is replaced by an annulus).  In some cases, one
of these sums can be calculated explicitly, thus simplifying
expressions and speeding up numerical computations (see an example in
\cite{Guillot19}).  The same argument is applicable to rectangles and
rectangular cuboids: $(a_1,b_1) \times (a_2,b_2) \times \ldots \times
(a_d,b_d) \subset \R^d$.  For instance, summation formulas were
intensively employed to study first-passage times for two particles
with reversible target-binding kinetics in a one-dimensional setting,
which can be mapped onto diffusion in a rectangle \cite{Grebenkov17}.

Rather than dwelling on mathematical proofs and technicalities, our
goal is to provide an intuitive practical guide over powerful
analytical tools for evaluating spectral sums that could be accessible
to non-experts.  While many common summation formulas often appearing
in applications are summarized, the review does not pretend to be
exhaustive.

\section{Expansion of meromorphic functions}
\label{sec:meromorphic}

The first strategy for evaluating spectral sums is based on the
corollary of the Mittag-Leffler theorem from complex analysis
\cite{Markushevich}.  This classical theorem describes the properties
of meromorphic functions, i.e., functions which are holomorphic (or
analytic) at all points except for a discrete set of isolated points
(their poles).  Any meromorphic function $F(z)$ can be represented as
a ratio of two holomorphic functions, $F(z) = f(z)/g(z)$, so that the
poles of $F(z)$ are just the zeros of $g(z)$.  While general theory
refers to an open subset of the complex plane $\C$, we restrict
discussion to meromorphic functions in $\C$, in which case holomorphic
functions are called entire functions (i.e., they can be represented
by a power series uniformly converging on compact sets).  Polynomials,
exponential function, sine and cosine, Bessel functions $J_n(z)$ of
the first kind of an integer order $n$, spherical Bessel functions
$j_n(z)$ of the first kind of an integer order $n$, their finite sums,
products and compositions, derivatives and integrals are examples of
entire functions.  In contrast, square root, logarithm and functions
with essential singularities (e.g., $e^{1/z}$) are not entire
functions.  The Mittag-Leffler theorem claims the existence of a
meromorphic function with prescribed poles.  As a consequence, any
meromorphic function can be decomposed as a sum of partial fractions.

In this light, a meromorphic function is a natural extension of
rational functions (i.e., ratio of two polynomials).  Any rational
function can be decomposed into a sum of partial fractions, e.g.,
\begin{equation}
\label{eq:example_rational}
\frac{(z - 1)^4}{z^2(z+1)} = z - 5 - \frac{5}{z} + \frac{1}{z^2} + \frac{16}{z+1} = 
z - 5 + \sum\limits_{k=1}^2 \sum\limits_{j=1}^{n_k} \frac{c_{k,j}}{(z - z_k)^j} \,,
\end{equation}
with the regular part $z-5$, the poles $z_1 = 0$ and $z_2 = -1$ of
order $n_1 = 2$ and $n_2 = 1$, and the weights $c_{1,1} = -5$,
$c_{1,2} = 1$ and $c_{2,1} = 16$.  The Mittag-Leffler theorem extends
such decompositions to meromorphic functions \cite{Markushevich}
(p. 297): any meromorphic function $F(z)$ with a prescribed sequence
of poles $\{z_k\}$ (with orders $n_k$) can be represented as
\begin{equation}
\label{eq:example_mero}
F(z) = \underbrace{F_0(z)}_{\textrm{regular part}} + \sum\limits_k \Biggl[
\underbrace{\sum\limits_{j=1}^{n_k} \frac{c_{k,j}}{(z - z_k)^j}}_{\textrm{principal parts}} \hskip 2mm 
+ \underbrace{P_k(z)}_{\textrm{regularization}}\Biggr] ,
\end{equation}
where the regular part $F_0(z)$ is an entire function, each principal
part is the sum of the partial fractions near the pole $z_k$ with
weights $c_{k,j}$, and $P_k(z)$ are polynomials.  The theorem states
that a meromorphic function is essentially determined by its principal
parts, up to an entire function.  Since a meromorphic function may
have infinite number of poles (i.e., the sum over $k$ can be
infinite), adding polynomials $P_k(z)$ to the principal parts in the
right-hand side of Eq. (\ref{eq:example_mero}) may be needed to ensure
the convergence of the series.  For instance, the decomposition of the
meromorphic function in Eq. (\ref{eq:example_rational}) includes the
entire function $F_0(z) = z-5$ and two principal parts.

The strategy for computing spectral sums consists in finding (or
guessing) an appropriate meromorphic function.  Let us first consider
spectral sums that can be reduced to 
\begin{equation}  \label{eq:series_general}
F(z) = \sum_k \frac{e_k}{z - z_k} \,,
\end{equation}
with a prescribed set of {\it simple} poles $\{z_k\}$ (i.e., $n_k =
1$) and coefficients $\{e_k\}$.  In the first step, one searches for
an entire function $g(z)$ whose complete set of zeros is the set
$\{z_k\}$.  In practice, this is a simple step because the eigenvalues
are often expressed in terms of zeros of some explicit entire
functions.  Next, one searches for another entire function $f(z)$ such
that $f(z_k) \ne 0$ for all $k$, and $F(z) = f(z)/g(z)$.  The zeros
$\{z_k\}$ are therefore the poles of the meromorphic function $F(z) =
f(z)/g(z)$.  Around $z \approx z_k$, one has $g(z) = g'(z_k)(z-z_k) +
O((z-z_k)^2)$, from which
\begin{equation}  \label{eq:auxil22}
F(z) = \frac{f(z)}{g(z)} = \frac{f(z_k)}{g'(z_k) (z-z_k)} + {\rm regular~part}  \qquad (z\to z_k),
\end{equation}
where prime denotes the derivative with respect to the argument of the
function.  The first term in (\ref{eq:auxil22}) is the principal part
of the function $F(z)$.  The function $f(z)$ should satisfy
$f(z_k)/g'(z_k) = e_k$ for each $k$.  Applying the corollary of the
Mittag-Leffler theorem, one has
\begin{equation}
\label{eq:meromorphic}
\frac{f(z)}{g(z)} = F(z) = F_0(z) + \sum\limits_k \left[\frac{f(z_k)}{g'(z_k)} ~ \frac{1}{z-z_k} + P_k(z)\right],
\end{equation}
where an entire function $F_0(z)$ (accounting for remaining regular
parts) and polynomials $P_k(z)$ (ensuring the convergence) have to be
found.  In many cases, the ``indefinite'' ingredients $F_0(z)$ and
$P_k(z)$ can be guessed or computed.  A practical strategy may consist
in computing numerically the truncated series in the right-hand side
and checking this relation (paying attention to the convergence which
may be critical for numerical analysis).

In the general case when a pole $z_k$ has the order $n_k$, the series
expansion around $z\approx z_k$ reads
\begin{eqnarray}  \nonumber
&& F(z) = \frac{f(z)}{g(z)} = \frac{b_0 + b_1 (z-z_k) + b_2 (z-z_k)^2 + \ldots}{(z-z_k)^{n_k} (a_0 + a_1 (z-z_k) + a_2 (z-z_k)^2 + \ldots)}  \\
&=& \frac{1}{(z-z_k)^{n_k}} \sum\limits_{m=0}^\infty c_m (z-z_k)^m = \sum\limits_{m=0}^{n_k-1} \frac{c_m}{(z-z_k)^{n_k - m}} + 
{\rm regular~part}, 
\end{eqnarray}
where
\begin{equation}
c_m = \frac{1}{a_0^{m+1}} \sum\limits_{i_0+\ldots+i_m + j = m} (-1)^{i_0} b_j a_0^{i_0} a_1^{i_1} \ldots a_m^{i_m} , 
\end{equation}
and the indices $i_0$, ... $i_m$, $j$ are nonnegative.  For instance,
one has
\begin{equation*}
c_0 = \frac{b_0}{a_0},  \quad
c_1 = \frac{b_0 a_1 - b_1 a_0}{a_0^2} , \quad
c_2 = \frac{b_0 a_1^2 + b_2 a_0^2 - b_1 a_1 a_0 - b_0 a_2 a_0}{a_0^3} , \ldots
\end{equation*}
The coefficients $a_m$ and $b_m$ are related to the derivatives of the
functions $f(z)$ and $g(z)$ at points $z_k$:
\begin{equation}
a_m = \frac{1}{(m+n_k)!} \left(\frac{d^{m+n_k} \, g(z)}{dz^{m+n_k}} \right)_{z = z_k} , \qquad
b_m = \frac{1}{m!} \left(\frac{d^m \, f(z)}{dz^m} \right)_{z = z_k} .
\end{equation}

In the particular case $e_k = n_k$, one formally gets an expansion of
the logarithmic derivative of $g(z)$:
\begin{equation}
\label{eq:decomp_entire}
\frac{g'(z)}{g(z)} = \sum\limits_k \frac{n_k}{z - z_k} \,.
\end{equation}
In a typical case of an infinite sequence of {\it simple} zeros (i.e.,
$n_k=1$)%
\footnote{
This mathematical result can be significantly extended as described by
\cite{Markushevich}.  Here we do not discuss convergence of the
series in the right-hand side of Eq. (\ref{eq:decomp_entire}),
especially since the most attention will be focused on the function
$\eta_n(z)$ defined by the convergent series in Eq. (\ref{eq:eta}).},
this relation can be made rigorous by requiring 
\begin{equation}  \label{eq:entire_cond}
\lim\limits_{k\to\infty} |z_k| = \infty  \qquad \mbox{and} \qquad
\sum\limits_k \frac{1}{|z_k|} = \infty .
\end{equation}
The relation (\ref{eq:decomp_entire}) can be formally obtained by
representing $g(z)$ as a product over its zeros,
\begin{equation*}
g(z) = g_0 \prod\limits_k (z - z_k)^{n_k}
\end{equation*}
(with a constant $g_0$), taking derivative and dividing it by $g(z)$.

\subsection{Basic example: interval with absorbing endpoints}

To illustrate this strategy, let us compute the Laplace-transformed
heat trace $\tilde{Z}(p)$ in the simple case of the Laplace operator
on the interval of length $R$ with Dirichlet boundary conditions at
both endpoints ($\LLambda = 0$), for which $\lambda_m =
\pi^2 m^2/R^2$, with $m = 1,2,3,\ldots$:
\begin{eqnarray}  \label{eq:example_ctan0}
\tilde{Z}(p) &=& \frac{R^2}{D} \sum\limits_{m=1}^\infty \frac{1}{pR^2/D + (\pi m)^2} 
= - \frac{R^2}{D} \sum\limits_{m=1}^\infty \frac{1}{z^2 - (\pi m)^2}    \\  \nonumber
&=& - \frac{R^2}{2zD} \sum\limits_{m=1}^\infty \left(\frac{1}{z - \pi m} + \frac{1}{z + \pi m}\right)
= \frac{R^2}{2zD} \left(\frac{1}{z} - \sum\limits_{k=-\infty}^\infty \frac{1}{z - \pi k} \right) ,
\end{eqnarray}
where we set $-z^2 = p R^2/D$.  One needs therefore to compute the
last sum with simple poles at $z_k = \pi k$ that are all the zeros of
the entire function $g(z) = \sin z$.  Applying
Eq. (\ref{eq:decomp_entire}), one gets 
\begin{equation} \label{eq:example_ctan1a}
\frac{\cos z}{\sin z} = \sum\limits_{k=-\infty}^\infty \frac{1}{z - \pi k} \,,
\end{equation}
which is still formal because the series in the right-hand side is
divergent.  To regularize this series, one can subtract $1/(\pi k)$
(for $k \ne 0$) whose contribution would be canceled anyway due to the
similar term for $-k$:
\begin{equation*}
\frac{\cos z}{\sin z} - \frac{1}{z} = \sum\limits_{k=-\infty \atop k\ne 0}^\infty \left(\frac{1}{z - \pi k} - \frac{1}{\pi k}\right)\,,
\end{equation*}
where the term with $k = 0$ was moved to the left-hand side.  The
subtracted term $1/(\pi k)$ is precisely the polynomial $P_k(z)$
(here, just a constant) in the general representation
(\ref{eq:example_mero}).  As the series in
Eq. (\ref{eq:example_ctan0}) is convergent, this regularization is
actually not needed, and the substitution of
Eq. (\ref{eq:example_ctan1a}) into Eq. (\ref{eq:example_ctan0}) yields
two equivalent classical summation formulas
\begin{subequations}  
\begin{align}  \label{eq:example_ctan2a}
\sum\limits_{k=1}^\infty \frac{1}{z^2 - \pi^2 k^2} & = \frac{\cos z}{2z \sin z} - \frac{1}{2z^2} \,, \\   \label{eq:example_ctan2b}
\sum\limits_{k=1}^\infty \frac{1}{z^2 + \pi^2 k^2} & = \frac{\cosh z}{2z \sinh z} - \frac{1}{2z^2} \,,
\end{align}
\end{subequations}
from which
\begin{equation}  \label{eq:example_ctan2}  
\tilde{Z}(p) = \frac{R \, \cosh (R\sqrt{p/D})}{2\sqrt{pD} \, \sinh(R\sqrt{p/D})} - \frac{1}{2p} \,.
\end{equation}
A similar derivation with the function $f(z) = 1$ in
(\ref{eq:meromorphic}) yields
\begin{equation}
\fl
\sum\limits_{k=1}^\infty \frac{(-1)^k}{z^2 - \pi^2 k^2} = \frac{1}{2z \sin z} - \frac{1}{2z^2} \,, \qquad
\sum\limits_{k=1}^\infty \frac{(-1)^k}{z^2 + \pi^2 k^2} = \frac{1}{2z \sinh z} - \frac{1}{2z^2} \,,
\end{equation}

The explicit form (\ref{eq:example_ctan2}) of the Laplace-transformed
heat trace $\tilde{Z}(p)$ facilitates the analysis of the short-time
and long-time asymptotic behaviors of the return-to-the-origin
probability discussed in Sec. \ref{sec:intro}.  The knowledge of
$\tilde{Z}(p)$ also allows one to compute some other spectral series,
in particular,
\begin{equation}
\sum\limits_{m=1}^\infty \frac{1}{\lambda_m^{n+1}} = \frac{(-1)^n \, D^{n+1}}{n!} 
\lim\limits_{p\to 0} \left(\frac{\partial^n}{\partial p^n} \tilde{Z}(p)\right)  \qquad (n=0,1,\ldots).
\end{equation}
For instance, one easily retrieves the even integer values of the
Riemann zeta-function,
\begin{equation}
\zeta(z) = \sum\limits_{m=1}^\infty \frac{1}{m^z} \,,
\end{equation}
as
\begin{equation}
\zeta(2n) = \pi^{2n} \frac{(-1)^{n-1}}{(n-1)!}  \lim\limits_{s\to 0} 
\frac{d^{n-1}}{ds^{n-1}} \left(\frac{\cosh (\sqrt{s})}{2\sqrt{s} \sinh (\sqrt{s})} - \frac{1}{s}\right)  \quad (n=1,2,\ldots) ,
\end{equation}
for instance
\begin{equation}  \label{eq:zeta}
\zeta(2) = \frac{\pi^2}{6}\,, \qquad  \zeta(4) = \frac{\pi^4}{90} \,, \qquad \ldots
\end{equation}
As shown below, many other spectral series can be reduced to
(\ref{eq:example_ctan2}).  While the same technique is applicable to a
more general case of an interval with partially absorbing endpoints
($\LLambda > 0$), we will discuss this case in
Sec. \ref{sec:1D_general}.

\subsection{Disk and ball}

In the case of a disk or a ball of radius $R$, the separation of
variables in polar or spherical coordinates leads again to an explicit
form of the eigenfunctions of the Laplace operator.  The eigenvalues
$\lambda_m$ can be written as $\alpha_{nk}^2/R^2$ with double indices
$n = 0,1,2,\ldots$ and $k = 1,2,\ldots$, where $\alpha_{nk}$ are
strictly positive simple zeros of an explicit entire function:
\begin{equation}
g_n(z) = \left\{ \begin{array}{l l} z \, J'_n(z) + h \, J_n(z) & \quad (d = 2), \\  
z \, j'_n(z) + h \, j_n(z) & \quad (d = 3), \\  \end{array}  \right.
\end{equation}
where $h = R/\ell$ is a nonnegative parameter (see
Table~\ref{tab:basic}).  In this case, the computation should be
performed separately for each $n$.  Note that the zero eigenvalue
appears only for Neumann boundary condition and will be treated
separately.

To illustrate this technique, we focus on computing the
Laplace-transformed heat kernel trace $\tilde{Z}(p)$, while more
sophisticated summation formulas will be reported in
Sec. \ref{sec:kernel} (see also Table \ref{tab:DS}).  For convenience,
we rewrite $\tilde{Z}(p)$ as
\begin{equation}
\tilde{Z}(p) = - \frac{R^2}{D} \sum\limits_n \sum\limits_{k=1}^\infty \frac{1}{-R^2p/D - \alpha_{nk}^2} 
= - \frac{R^2}{D} \sum\limits_n \eta_n\bigl(iR\sqrt{p/D}\bigr) ,
\end{equation}
where 
\begin{equation}
\label{eq:eta}
\eta_n(z) = \sum\limits_{k=1}^\infty \frac{1}{z^2 - \alpha_{nk}^2} = \frac{1}{2z} \sum\limits_{k=1}^\infty 
\left(\frac{1}{z - \alpha_{nk}} + \frac{1}{z + \alpha_{nk}}\right) \,,
\end{equation}
and $z = R\sqrt{p/D}$.  According to the form (\ref{eq:Um}) of
Laplacian eigenfunctions, the function $\eta_n(z)$ can be associated
with the $n$-th Fourier (or spherical harmonic) mode.
Note that $-\alpha_{nk}$ are also zeros of the function $g_n(z)$ due
to its symmetry.  Moreover, the function $g_n(z)$ vanishes at $z = 0$
for any $n > 0$ but also for $n = 0$ in the case of Neumann boundary
condition.  Since the set $\{0, \pm \alpha_{nk}\}$ contains {\it all}
zeros of $g_n(z)$, one can apply the corollary of the Mittag-Leffler
theorem to get
\begin{equation}
\label{eq:eta_1}
\eta_n(z) = \frac{1}{2z} \frac{g'_n(z)}{g_n(z)} - \frac{n}{2z^2} \,,
\end{equation}
where the second term accounts for the zero at $0$, which is excluded
from $\eta_n(z)$.
For Neumann boundary condition, $z=0$ is also a simple zero of
$g_0(z)$, and as it was accounted twice, it should be subtracted:
\begin{equation}
\label{eq:eta2}
\eta_0(z) = \frac{1}{2z} \frac{g'_0(z)}{g_0(z)} - \frac{1}{2z^2} \, .
\end{equation}
Note that the condition (\ref{eq:entire_cond}) is satisfied thanks to
the asymptotic behavior $\alpha_{nk}\propto k$ (for fixed $n$).
As shown in \cite{Grebenkov07b}, $\eta_n(z)$ from (\ref{eq:eta_1})
with $n>0$ can also be written for a ball in any dimension and
boundary condition as
\begin{equation}
\label{eq:etan_general}
\eta_n(z) = \frac{(h-n-d+2) - \bigl[z^2+(2n+d-2)(h-n-d+2)\bigr] \psi_{n,d}(z)}{2z^2\bigl[1+(h-n-d+2)\psi_{n,d}(z)\bigr]} \,,
\end{equation}
where 
the function $\psi_{n,d}(z)$ is defined by the recurrent relation
\begin{equation}
\psi_{n+1,d}(z) = \frac{2n+d-2}{z^2} - \frac{1}{z^2~\psi_{n,d}(z)}  \qquad (n \geq 1),
\end{equation}
and
\begin{equation}
\psi_{1,d}(z) = \left\{ \begin{array}{l l} 
\displaystyle \frac{\sin z}{z \,\cos z} & \quad (d=1),   \vspace*{1mm} \\  
\displaystyle \frac{J_1(z)}{z \,J_0(z)}   & \quad (d=2),  \\
\displaystyle \frac{j_1(z)}{z \, j_0(z)}   & \quad (d=3).  \\ \end{array} \right.
\end{equation}
Note that the above representation with $d = 1$ is also valid for a
symmetric interval $(-R,R)$, see details in \cite{Grebenkov07b}.

As an example, let us consider a disk with Dirichlet boundary
condition ($\LLambda = 0$), for which Eq. (\ref{eq:eta_1}) implies the
summation formula:
\begin{equation}
\sum\limits_{k=1}^\infty \frac{1}{z^2 - \alpha_{nk}^2} = \frac{1}{2z} \frac{J'_n(z)}{J_n(z)} 
= - \frac{1}{2z} \frac{J_{n+1}(z)}{J_n(z)} \,.
\end{equation}
This is a particular case of the summation formula (see \cite{Watson},
p. 498)
\begin{equation}  \label{eq:Bessel_ident}
\frac{J_{\nu+1}(z)}{J_\nu(z)} = \sum\limits_{k=1}^\infty \frac{2z}{\alpha_{\nu k}^2 - z^2}    \qquad (\nu > -1),
\end{equation}
where $\pm \alpha_{\nu k}$ are zeros of $z^{-\nu} J_\nu(z)$.
Evaluating the derivatives of this identity at $z = 0$, Rayleigh 
found other summation formulas \cite{Rayleigh74}, e.g.,
\begin{equation}  \label{eq:sum_alpha}
\sum\limits_{k=1}^\infty \frac{1}{\alpha_{\nu k}^{2}} = \frac{1}{2^2(\nu+1)} \,,  \qquad 
\sum\limits_{k=1}^\infty \frac{1}{\alpha_{\nu k}^{4}} = \frac{1}{2^4(\nu+1)^2 (\nu+2)}\,, \qquad  \cdots
\end{equation}
(see also \cite{Watson,Sneddon60,Giusti16}).  At $\nu = 1/2$,
$z^{-\nu} J_\nu(z) = \sqrt{2/\pi} \, \frac{\sin z}{z}$ so that
Eqs. (\ref{eq:Bessel_ident}, \ref{eq:sum_alpha}) are reduced to
Eqs. (\ref{eq:example_ctan0}, \ref{eq:zeta}) for the interval.
Similarly, one gets a representation of the inverse of the Bessel
function as
\begin{equation}
\frac{1}{J_n(z)} = \PP_n(1/z) + \left\{ \begin{array}{l l}
\displaystyle 2z \sum\limits_{k=1}^\infty \frac{1}{J'_n(\alpha_{nk}) (z^2 - \alpha_{nk}^2)} & (n~\mbox{is odd}), \\
\displaystyle  2 \sum\limits_{k=1}^\infty \frac{\alpha_{nk}}{J'_n(\alpha_{nk}) (z^2 - \alpha_{nk}^2)} & (n~\mbox{is even}), \end{array}  \right. 
\end{equation}
where $\PP_n$ is the polynomial of degree $n$ that accounts for the
pole at $0$:
\begin{equation}
\PP_0(x) = 0, \quad \PP_1(x) = 2x, \quad \PP_2(x) = 8x^2, \quad \PP_3(x) = 3x + 48x^3, \quad \ldots
\end{equation}
The explicit form of this polynomial can be obtained by expanding
$z^n/J_n(z)$ in a Taylor series around $0$ and keeping the terms up to
the $n$-th order.

Once $\eta_n(z)$ is known, one can establish a number of useful
relations for more complicated sums involving the Laplacian
eigenvalues.  Such spectral sums naturally appear when the quantity of
interest depends on several time variables, $t_1, \ldots, t_j$, as,
e.g., the $j$-point correlation function in (\ref{eq:Xnt}).  In this
case, the Laplace transform is applied to each time variable, and the
Laplace-transformed quantity in (\ref{eq:Xnp}) depends on multiple
conjugate parameters $p_1,\ldots,p_j$.
First, one can compute the sums
\begin{equation}
\label{eq:eta_useful2a}
\eta_n^{(j)}(z_1,\ldots,z_j) \equiv \sum\limits_{k=1}^\infty \prod\limits_{i=1}^j
\frac{1}{z^2_i - \alpha_{nk}^2} 
\end{equation}
iteratively, applying the identity
\begin{equation}
\label{eq:eta_useful2b}
\eta_n^{(j)}(z_1,\ldots,z_j) = \frac{\eta_n^{(j-1)}(z_1,\ldots,z_{j-2},z_{j-1}) - 
\eta_n^{(j-1)}(z_1,\ldots,z_{j-2},z_j)}{z^2_j - z^2_{j-1}} \,,
\end{equation}
with $\eta_n^{(1)}(z) \equiv \eta_n(z)$.  In particular,
\begin{eqnarray}
\label{eq:Theta_n}
\fl
\eta^{(2)}_n(z_1,z_2) &=& \frac{\eta_n(z_1) - \eta_n(z_2)}{z_2 - z_1} , \\
\fl
\eta^{(3)}_n(z_1,z_2,z_3) &=& \frac{1}{z^2_3-z^2_2}\left(\frac{\eta_n(z_1)-\eta_n(z_2)}{z^2_2-z^2_1} 
- \frac{\eta_n(z_1)-\eta_n(z_3)}{z^2_3-z^2_1}\right) , \\  \nonumber
\fl
\eta^{(4)}_n(z_1,z_2,z_3,z_4) &=&
\frac{1}{z^2_4-z^2_3}\left(\frac{1}{z^2_3-z^2_2}\left[\frac{\eta_n(z_1)-\eta_n(z_2)}{z^2_2-z^2_1} 
- \frac{\eta_n(z_1)-\eta_n(z_3)}{z^2_3-z^2_1}\right]  \right. \\  
&&  \left. - \frac{1}{z^2_4-z^2_2}\left[\frac{\eta_n(z_1)-\eta_n(z_2)}{z^2_2-z^2_1} - 
\frac{\eta_n(z_1)-\eta_n(z_4)}{z^2_4-z^2_1}\right]\right) . 
\end{eqnarray}
Multiple differentiation of Eq.~(\ref{eq:eta_useful2a}) with respect
to the variables $s_i = z^2_i$ further extends the set of useful
relations for any positive integers $\sigma_1$, ..., $\sigma_j$:
\begin{equation}
\label{eq:eta_useful3}
\sum\limits_{k=1}^\infty \prod\limits_{i=1}^j \frac{1}{(s_i - \alpha_{nk}^2)^{\sigma_i+1}} = 
 \frac{(-1)^{\sigma_1+\ldots+\sigma_j}}{\sigma_1! \ldots \sigma_j!} ~ 
\frac{\partial^{\sigma_1+\ldots+\sigma_j}}{\partial s_1^{\sigma_1} \ldots \partial s_j^{\sigma_j}}
~ \eta_n^{(j)}(\sqrt{s}_1,\ldots,\sqrt{s}_j) .
\end{equation}
For instance, for $j=1$ and $\sigma = 0,1,2,\ldots$, one gets
\begin{equation}
\label{eq:eta_useful4}
\sum\limits_{k=1}^\infty \frac{1}{(s - \alpha_{nk}^2)^{\sigma+1}} = 
 \frac{(-1)^{\sigma}}{\sigma!} ~ \frac{\partial^{\sigma} \eta_n(\sqrt{s})}{\partial s^{\sigma}} \,,
\end{equation}
from which the sum of inverse powers of eigenvalues (as in
Eq. (\ref{eq:sum_alpha})) can be evaluated by setting $z\to 0$ (or
$s\to 0$).   For instance, setting $\sigma = 0$, one retrieves
(see, e.g., \cite{Ahmed83} and references therein):
\begin{equation}  \label{eq:alpha_2sum}
\sum\limits_{k=1}^\infty \frac{1}{\alpha_{nk}^2} = \frac{n+h+2}{4(n+1)(n+h)} \,.
\end{equation}
Although cumbersome, these expressions allow one to
rigorously compute many spectral sums and related quantities involving
the Laplacian eigenvalues (see \cite{Grebenkov07b,Grebenkov11} for
examples and applications).

As an example, we discuss the Calogero's formula for the zeros
$\alpha_{\nu k}$ of the function $g_n(z) = z J'_\nu(z) + h J_\nu(z)$
with any real $\nu > - 1$ \cite{Calogero77,Ahmed78a}
\begin{equation} \label{eq:Calogero} 
\sum\limits_{k=1 \atop k\ne j}^\infty \frac{1}{\alpha_{\nu j}^2 - \alpha_{\nu k}^2} 
= \frac{1}{2(\alpha_{\nu j}^2 + h^2 - \nu^2)} - \frac{\nu+1}{2\alpha_{\nu j}^2} \,.
\end{equation}
This formula can be easily derived for integer $\nu = n$ by evaluating
the limit:
\begin{equation}
\sum\limits_{k=1 \atop k\ne j}^\infty \frac{1}{\alpha_{nj}^2 - \alpha_{nk}^2}
= \lim\limits_{\ve\to 0} \left(\eta_n(\alpha_{nj} + \ve) - \frac{1}{(\alpha_{nj}+\ve)^2 - \alpha_{nj}^2}\right).
\end{equation}
Using the explicit form (\ref{eq:eta_1}) of the function $\eta_n(z)$,
one can compute the limit by expanding $g_n(z)$ and $g'_n(z)$ in a
Taylor series around $z = \alpha_{nj}$.  After simplifications, one
gets
\begin{equation}
\sum\limits_{k=1 \atop k\ne j}^\infty \frac{1}{\alpha_{nj}^2 - \alpha_{nk}^2}
= \frac{g''_n(\alpha_{nj})}{4\alpha_{nj} g'_n(\alpha_{nj})} - \frac{n+1/2}{2\alpha_{nj}^2} \,.
\end{equation}
Using the Bessel equation, 
\begin{equation*}
z^2 J_n''(z) + z J_n'(z) + (z^2 - n^2) J_n(z) = 0,
\end{equation*}
one can express in two dimensions
\begin{eqnarray}
g'_n(z) &=& h J'_n(z) + \frac{n^2 - z^2}{z} J_n(z), \\
g''_n(z) &=& \frac{n^2 - z^2 - h}{z} J'_n(z) + \frac{h(n^2 - z^2) - n^2 - z^2}{z^2} J_n(z) .
\end{eqnarray}
Since $\alpha_{nj}$ is a zero of $g_n(z)$, one finally gets
\begin{equation}
\frac{g''_n(\alpha_{nj})}{g'_n(\alpha_{nj})} = \frac{\alpha_{nj}^2 + n^2 - h^2}{\alpha_{nj}(\alpha_{nj}^2 + h^2 - n^2)} \,,
\end{equation}
from which the Calogero's formula (\ref{eq:Calogero}) follows
for an integer $\nu = n$.
%
Similarly, one can deduce the summation formulas for higher-order
inverse differences between eigenvalues, see
\cite{Calogero77b,Ahmed78b,Ahmed83,Baricz15}.  Interestingly, Calogero
and co-workers considered (\ref{eq:Calogero}) as an infinite system of
nonlinear equations that determine the zeros $\alpha_{\nu j}$,
together with the normalization relation (\ref{eq:alpha_2sum})
extended to non-integer $\nu$.

In three dimensions, one uses the modified Bessel equation to express
\begin{eqnarray}
\fl
g'_n(z) &=& (h-1) j'_n(z) + \frac{n(n+1) - z^2}{z} j_n(z), \\
\fl
g''_n(z) &=& \frac{n(n+1) - z^2 - 2(h-1)}{z} j'_n(z) + \frac{(h-2)n(n+1) - hz^2}{z^2} j_n(z) ,
\end{eqnarray}
from which
\begin{equation}
\frac{g''_n(\alpha_{nj})}{g'_n(\alpha_{nj})} = 2 \frac{n(n+1) - h(h-1)}{\alpha_{nj}(\alpha_{nj}^2 + h(h-1) - n(n+1))} \,,
\end{equation}
and thus
\begin{equation}
\sum\limits_{k=1 \atop k\ne j}^\infty \frac{1}{\alpha_{nj}^2 - \alpha_{nk}^2}
= \frac{1}{2(\alpha_{nj}^2 + h(h-1) - n(n+1))} - \frac{n+3/2}{2\alpha_{nj}^2} \,.
\end{equation}  
This is an extension of the Calogero's formula to the zeros
$\alpha_{nk}$ of the function $g_n(z) = z j'_n(z) + h j_n(z)$
involving spherical Bessel function and its derivative.

As the review is focused on the eigenvalues of the Laplace operator in
rotation-invariant domains (here, a disk or a ball), we restricted the
discussion to Bessel function $J_n(z)$ and spherical Bessel function
$j_n(z)$ of integer index $n$.  However, as illustrated above, many
summation formulas of this section are valid if $n$ is replaced by a
real index $\nu > -1$.  In this more general setting, one needs to
consider $z^{-\nu} J_\nu(z)$ or $z^{-\nu} j_\nu(z)$, which are again
entire functions.  Note that the zeros of Bessel and spherical Bessel
functions with non-integer indices are naturally related to the
Laplacian eigenvalues for circular and spherical sectors, respectively
\cite{Carslaw}.  In spite of their practical interest, we do not
discuss the summation formulas for these domains.  Finally, for
$d$-dimensional balls with $d > 3$, one needs to consider the zeros of
ultraspherical Bessel functions $z^{1-d/2} J_{d/2-1+n}(z)$.

\begin{table}
\begin{center}
\begin{tabular}{|c|c|c|}  \hline
         &  Disk  &  Ball \\  \hline
Domain $\Omega$ &  $\{\x\in\R^2 ~:~ |\x| < R\}$ &  $\{\x\in\R^3 ~:~ |\x| < R\}$ \\
Coordinates &  Polar $(r,\phi)$           & Spherical $(r,\theta,\phi)$ \\
Laplacian $\Delta$  & $\partial_r^2 + \frac{1}{r}\, \partial_r$ &  
$\partial_r^2 + \frac{2}{r} \partial_r$ \\
& $+ \frac{1}{r^2}\, \partial^2_\phi$ & $+ \frac{1}{r^2} \biggl(\frac{1}{\sin\theta} \partial_\theta \sin\theta \partial_\theta +
\frac{1}{\sin^2\theta}\partial^2_\phi\biggr)$ \\
eigenfunctions $U$ &  $C_{nk} \, J_n(\alpha_{nk} r/R)\,  e^{in\phi}$ & $C_{nk} \, j_n(\alpha_{nk} r/R) \, P_n^l(\cos\theta) \, e^{il\phi}$ \\
indices  &  $n\in \Z$       & $n = 0,1,2,\ldots$,~~ $l = -n,\ldots,n$ \\
function $g_n(z)$   &  $zJ'_n(z) + h J_n(z)$     &  $zj'_n(z) + h j_n(z)$ \\
eigenvalues   & $\lambda_{nk} = \alpha_{nk}^2/R^2$ & $\lambda_{nk} = \alpha_{nk}^2/R^2$  \\
$C_{nk}$ for $\LLambda > 0$ & $\frac{\beta_{nk}}{\sqrt{\pi}\, J_n(\alpha_{nk}) R}$  & 
  $\frac{\sqrt{2n+1}\, \beta_{nk}}{\sqrt{2\pi}\, j_n(\alpha_{nk}) R^{3/2}}$ \\ 
$C_{nk}$ for $\LLambda = 0$ & $\frac{1}{\sqrt{\pi}\, J'_n(\alpha_{nk}) R}$  & 
  $\frac{\sqrt{2n+1}}{\sqrt{2\pi}\, j'_n(\alpha_{nk}) R^{3/2}}$ \\ 
coefficient $\beta_{nk}$ & $\frac{\alpha_{nk}}{\sqrt{\alpha_{nk}^2 - n^2 + h^2}}$ &
  $\frac{\alpha_{nk}}{\sqrt{\alpha_{nk}^2 - n(n+1) + h(h-1)}}$ \\    \hline
\end{tabular}
\end{center}
\caption{
Laplacian eigenfunctions and eigenvalues in a disk and a ball of
radius $R$.  We recall that $J_n(z)$ and $j_n(z)$ are the Bessel and
spherical Bessel functions of the first kind, $P_n^l(z)$ are the
associated Legendre polynomials, and $h = R/\LLambda$.  Here
$\alpha_{nk}$ are all {\it strictly} positive zeros of the function
$g_n(z)$, enumerated by $k = 1,2,3,\ldots$, except for the Neumann
boundary condition ($h = 0$), for which $\alpha_{01} = 0$ and
$\beta_{01} = 1$ for the disk, as well as $\alpha_{01} = 0$ and
$\beta_{01} = \sqrt{3/2}$ for the ball.  Note that the eigenfunctions
for the disk are often written in the real-valued form, in which
$e^{in\phi}$ is replaced by $\cos(n\phi)$ and $\sin(n\phi)$ with $n =
0,1,2,\ldots$.}
\label{tab:basic}
\end{table}

\section{Fourier-Bessel and Dini series}
\label{sec:Dini}

The second strategy for evaluating spectral sums relies on
Fourier-Bessel and Dini series which extend Fourier series to Bessel
functions.  The Fourier-Bessel series involve the zeros of a Bessel
function $J_\nu(z)$ or of its derivative $J'_\nu(z)$, whereas Dini
series involve the zeros of a linear combination $z J'_\nu(z) + h
J_\nu(z)$ with a constant $h$.  As these zeros naturally appear from
Dirichlet, Neumann and Robin boundary conditions on Laplacian
eigenfunctions (see Table \ref{tab:basic}), we refer to them as
Dirichlet, Neumann and Robin cases.

A continuous function $f(x)$ on $(0,1)$, integrable on $(0,1)$ with
the weight $x^{1/2}$ and of limited total variation, admits
\cite{Watson,Sneddon60}:
\begin{itemize}
\item the Fourier-Bessel expansion (for $\nu \geq -1/2$)
\begin{equation}  \label{eq:FB_Dirichlet}
f(x) = 2\sum\limits_{k=1}^\infty \frac{J_\nu(\alpha_{\nu k}
x)}{J_{\nu+1}^2(\alpha_{\nu k})} \int\limits_0^1 dy \, y\, f(y) \,
J_\nu(\alpha_{\nu k} y),
\end{equation}
where $\alpha_{\nu k}$ are the positive zeros of $J_\nu(z)$ (Dirichlet case);  

\item the Fourier-Bessel expansion (for $\nu > 0$)
\begin{equation}  \label{eq:FB_Neumann}
f(x) = 2\sum\limits_{k=1}^\infty \frac{J_\nu(\alpha_{\nu k} x)}{J_{\nu}^2(\alpha_{\nu k})} \frac{\alpha_{\nu k}^2}{\alpha_{\nu k}^2-\nu^2}
 \int\limits_0^1 dy \, y\, f(y) \, J_\nu(\alpha_{\nu k} y) ,
\end{equation}
where $\alpha_{\nu k}$ are the positive zeros of $J'_\nu(z)$ (Neumann
case);

\item the Dini expansion (for $\nu \geq -1/2$ and $\nu + h > 0$)
\begin{equation}  \label{eq:Dini}
f(x) = 2\sum\limits_{k=1}^\infty \frac{J_\nu(\alpha_{\nu k} x)}{J_{\nu}^2(\alpha_{\nu k})} \frac{\alpha_{\nu k}^2}{\alpha_{\nu k}-\nu^2+h^2}
 \int\limits_0^1 dy \, y\, f(y) \, J_\nu(\alpha_{\nu k} y) ,
\end{equation}
where $\alpha_{\nu k}$ are the positive zeros of $z J'_\nu(z) + h
J_\nu(z)$ (Robin case).  
\end{itemize}
Note that the continuity assumption and some other constraints can be
weakened (see details in \cite{Watson}).  If the Fourier series can be
seen as spectral expansions on over Laplacian eigenfunctions on an
interval, Fourier-Bessel and Dini series are essentially the spectral
expansions over radial parts of the Laplacian eigenfunctions on the
disk (see also Sec. \ref{sec:kernel}).

As pointed out by Watson \cite{Watson}, the coefficients of these
expansions, given by integrals with the function $f(y)$, are rarely
known explicitly.  Among such examples, one has
\begin{equation}
x^\nu = \sum\limits_{k=1}^\infty \frac{2J_\nu(\alpha_{\nu k}x)}{\alpha_{\nu k} \, J_{\nu+1}(\alpha_{\nu k})}  \qquad (0 \leq x < 1)
\end{equation}
for $\alpha_{\nu k}$ being zeros of $J_\nu(z)$, and 
\begin{equation}
x^\nu = \sum\limits_{k=1}^\infty \frac{2 \alpha_{\nu k} \, J_{\nu+1}(\alpha_{\nu k}) \, J_\nu(\alpha_{\nu k}x)}
{(\alpha_{\nu k}^2 - \nu^2 + h^2) [J_\nu(\alpha_{\nu k})]^2}  \qquad (0 \leq x \leq 1)
\end{equation}
for $\alpha_{\nu k}$ being zeros of $zJ'_\nu(z) + h J_\nu(z)$ (under
the additional condition $\nu + h > 0$).

Relying on these expansions, Sneddon derived several kinds of
summation formulas for some positive integer $m$
\cite{Sneddon60}
\begin{itemize}
\item Dirichlet case: $\alpha_{\nu k}$ are positive zeros of $J_\nu(z)$
\begin{equation}  \label{eq:S_D1}
\sum\limits_{k=1}^\infty \frac{1}{\alpha_{\nu k}^{2m}} =
\left\{ \begin{array}{l l}  \frac{1}{4(\nu+1)}  &  (m = 1), \\  
\frac{1}{16(\nu+1)^2 (\nu+2)} & (m=2), \\  
\frac{1}{32(\nu+1)^3 (\nu+2) (\nu+3)} & (m=3), \\
\frac{5\nu+11}{256(\nu+1)^4 (\nu+2)^2 (\nu+3) (\nu+4)} & (m=4) \\
\end{array}  \right.
\end{equation}
and
\begin{equation}
\sum\limits_{k=1}^\infty \frac{1}{\alpha_{\nu k}^{2m-\nu+1} J_{\nu+1}(\alpha_{\nu k})} = 
\left\{ \begin{array}{l l} \frac{ 2^{\nu-3} \, \Gamma(\nu+1)}{(\nu+1)} & (m=1), \\
\frac{2^{\nu-6} (\nu+3) \Gamma(\nu+1)}{(\nu+1)^2 (\nu+2)} & (m=2), \\
\frac{2^{\nu-8} (\nu^2+8\nu+19) \Gamma(\nu+1)}{3 (\nu+1)^3 (\nu+2)(\nu+3)} & (m=3). \\
\end{array}  \right.
\end{equation}
(note that (\ref{eq:S_D1}) reproduces Rayleigh's formulas
(\ref{eq:sum_alpha})).

\item Neumann case: $\alpha_{\nu k}$ are positive zeros of $J'_\nu(z)$
\begin{equation}
\sum\limits_{k=1}^\infty \frac{1}{\alpha_{\nu k}^{2m} (\alpha_{\nu k}^2 - \nu^2)}  =
\left\{ \begin{array}{l l}  \frac{1}{4\nu^2(\nu+1)}  &  (m = 1), \\  
\frac{3\nu+4}{16\nu^3(\nu+1)^2 (\nu+2)} & (m=2), \\  
\frac{5\nu^2+16\nu+12}{32\nu^4(\nu+1)^2 (\nu+2) (\nu+3)} & (m=3), \\
\frac{35\nu^4+273\nu^3+768\nu^2+896\nu+392}{256\nu^5(\nu+1)^4 (\nu+2)^2 (\nu+3) (\nu+4)} & (m=4) \\
\end{array}  \right.
\end{equation}
and
\begin{equation}
\sum\limits_{k=1}^\infty \frac{1}{\alpha_{\nu k}^{2m-\nu} (\alpha_{\nu k}^2 - \nu^2) J_{\nu}(\alpha_{\nu k})} = 
\left\{ \begin{array}{l l} \frac{1}{8\nu(\nu+1)} & (m=1), \\
\frac{\nu^2+7\nu+8}{64\nu^2(\nu+1)^2 (\nu+2)} & (m=2), \\  
\frac{\nu^4+14\nu^3+91\nu^2+210\nu+144}{768\nu^3(\nu+1)^3 (\nu+2) (\nu+3)} & (m=3). \\
\end{array}  \right.
\end{equation}

\item Robin case: $\alpha_{\nu k}$ are positive zeros of $z J'_\nu(z) + h J_\nu(z)$
\begin{equation}  \label{eq:Sneddon_S}
\sum\limits_{k=1}^\infty \frac{1}{\alpha_{\nu k}^{2m} (\alpha_{\nu k}^2 - \nu^2 + h^2)} =
\left\{ \begin{array}{l l}  \frac{1}{2(h + \nu)} &  (m = 0), \\  
\frac{1}{4(h + \nu)^2(\nu+1)}  & (m=1), \\  
\frac{h + 3\nu + 4}{16 (h + \nu)^3 (\nu+1)^2 (\nu+2)}  & (m=2), \\
\frac{h^2 + 2(2\nu+3)h + 5\nu^2 + 16\nu + 2}{32 (h + \nu)^4 (\nu+1)^3 (\nu+2)(\nu+3)}  & (m=3). \\
\end{array}  \right.
\end{equation}

\end{itemize}
Note that the Sneddon's technique can be used to derive such formulas
for larger values of $m$.

%
Sneddon also gave some examples of other series, e.g.,
\begin{equation}
\sum\limits_{k=1}^\infty \frac{\sin \alpha_{0k}}{\alpha_{0k}^2 J_1(\alpha_{0k})} = \frac12 \,, \qquad
\sum\limits_{k=1}^\infty \frac{\sin \alpha_{0k}}{\alpha_{0k}^3 J_1^2(\alpha_{0k})} = \frac{1 - \ln 2}{2} 
\end{equation}
for the Dirichlet case, and
\begin{equation}
\sum\limits_{k=1}^\infty \frac{\sin \alpha_{0k}}{(\alpha_{0k}^2 + h^2) J_0(\alpha_{0k})} = \frac{1}{2h}
\end{equation}
for the Robin case.

In summary, Fourier-Bessel and Dini series present a powerful tool for
deriving spectral sums over the zeros of Bessel functions.  However,
as such series involve only Bessel functions of the first kind,
$J_\nu(z)$, they are not directly applicable to circular annuli or
spherical shells, for which the Laplacian eigenfunctions also include
Bessel functions of the second kind, $Y_\nu(z)$.  In the next section,
we discuss a different method based on the direct computation of the
Laplace-transformed heat kernel.  Here, the radial coordinate will be
restricted on an interval $(a,b)$, and the Sturm-Liouville theory will
ensure the completeness of a basis formed by linear combinations of
$J_\nu(z)$ and $Y_\nu(z)$.

\section{Laplace-transformed heat kernel}
\label{sec:kernel}

The two strategies discussed in the previous sections are well adapted
for evaluating spectral sums involving the eigenvalues of the Laplace
operator on an interval, a disk, and a ball.  In particular, the
associated eigenfunctions are expressed through entire functions:
sine, cosine, Bessel and spherical Bessel functions of the first kind.
However, these strategies are not suitable for circular annuli and
spherical shells because Bessel and spherical Bessel functions of the
second kind, $Y_n(z)$ and $y_n(z)$, are not entire functions.
Similarly, Fourier-Bessel and Dini series do not operate with $Y_n(z)$
and $y_n(z)$.  For this reason, we discuss here in more detail the
third strategy for evaluating spectral sums that is also applicable to
the Laplace operator in circular annuli and spherical shells.  In
these rotation-invariant domains, the separation of variables reduces
the computation of the Laplacian eigenvalues to the analysis of radial
second-order differential operators $\L_d$ in Eq. (\ref{eq:L_radial}).

\subsection{General case}

Let us first recall the basic elements of the Sturm-Liouville theory
can be found in most textbooks on differential equations and
mathematical physics, e.g., \cite{Stakgold,Boyce}.
Let $\L$ be a self-adjoint second-order differential operator on an
interval $(a,b)$, with prescribed boundary conditions at endpoints $a$
and $b$.  As we will focus on the particular well-studied case of the
operators $\L_d$ in Eq. (\ref{eq:L_radial}), we skip mathematical
details and eventual constraints known from the spectral theory of
ordinary differential operators \cite{Birman}.  In particular, we
postulate that the spectrum of the operator $\L$ is discrete that is
clearly the case in our setting. 

The third strategy relies on computing the Green's function
$G_q(r,r_0)$, i.e., the resolvent of the operator $\L$ (acting on $r$)
that obeys the equation
\begin{equation}
(q^2 - \L) G_q(r,r_0) \omega(r_0) = \delta(r - r_0) ,
\end{equation}
where $\omega(r_0)$ is a given positive weighting function (see
below).  On one hand, the standard construction of $G_q(r,r_0)$
involves two linearly independent solutions $v_q^a$ and $v_q^b$ of the
homogeneous equation $(q^2 - \L) v_q^{a,b} = 0$ such that $v_q^a$
satisfies the boundary condition at $r=a$ and $v_q^b$ satisfies the
boundary condition at $r=b$.  From these two solutions, one can
construct the Green's function as
\begin{equation}
G_q(r,r_0) = \omega^{-1}(r_0) \times \left\{ \begin{array}{c} A v_q^a(r)  \quad (a < r < r_0), \\  B v_q^b(r) \quad (r_0 < r < b), \\ \end{array} \right.
\end{equation}
where the coefficients $A$ and $B$ are determined by matching these
parts at the point $r = r_0$ to ensure the continuity of $G_q(r,r_0)$
and the unit jump of its derivative.  One gets thus
\begin{equation}  \label{eq:general_G1}
\fl
G_q(r,r_0) = \frac{\omega^{-1}(r_0)}{v_q^b(r_0) v_q^{a\,\prime}(r_0) - v_q^a(r_0) v_q^{b\,\prime}(r_0)} \times
\left\{ \begin{array}{c} v_q^b(r_0) v_q^a(r)  \quad (a \leq r \leq r_0), \\  v_q^a(r_0) v_q^b(r) \quad (r_0 \leq r \leq b). \\ \end{array} \right. 
\end{equation}

On the other hand, the function $v_{i\sqrt{\lambda}}^a(r)$ with a
nonnegative $\lambda$ is a natural candidate to be an eigenfunction of
the operator $\L$ (a similar construction is applicable to
$v_{i\sqrt{\lambda}}^b(r)$).  Indeed, this function obeys the
eigenvalue equation, $\L u + \lambda u = 0$, with the imposed boundary
condition at $r = a$.  The remaining boundary condition at $r = b$
determines the (infinite) set of eigenvalues of this operator that we
denote as $\lambda_k$, with the index $k = 1,2,\ldots$.  The
corresponding eigenfunctions are then
\begin{equation}  \label{eq:general_eigen}
u_k(r) = i v_{i\sqrt{\lambda_k}}^a (r),
\end{equation}
where the prefactor $i$ was introduced for convenience.  Denoting
\begin{equation}
c_k^2 = \left(\int\limits_a^b dr \, \omega(r)\, |u_k(r)|^2 \right)^{-1}
\end{equation}
the $L_2$-normalization coefficient of $u_k$ with the weighting
function $\omega(r)$, one can express the Green's function via the
spectral decomposition over the eigenfunctions of the operator $\L$
\begin{equation}  \label{eq:general_G2}
G_q(r,r_0) = \sum\limits_{k=1}^\infty c_k^2 \frac{u_k(r) \, u_k^*(r_0)}{q^2 + \lambda_k} \,.
\end{equation}
Equating relations (\ref{eq:general_G1}, \ref{eq:general_G2}), one
gets an explicit summation formula over the eigenvalues $\lambda_k$:
\begin{equation}  \label{eq:general0}
\fl
\sum\limits_{k=1}^\infty c_k^2 \frac{u_k(r) \, u_k^*(r_0)}{q^2 + \lambda_k} =
\frac{\omega^{-1}(x_0)}{v_q^b(r_0) v_q^{a\,\prime}(r_0) - v_q^a(r_0) v_q^{b\,\prime}(r_0)} \times
\left\{ \begin{array}{c} v_q^b(r_0) v_q^a(r)  \quad (a \leq r \leq r_0), \\  v_q^a(r_0) v_q^b(r) \quad (r_0 \leq r \leq b). \\ \end{array} \right.
\end{equation}
Multiplying both sides of this formula by an integrable function
$f(r,r_0)$ (with the weighting function $\sqrt{\omega(r)}$) and
integrating over $r$ and $r_0$, one can deduce many other summation
formulas.

In the following, we focus on Robin boundary condition
(\ref{eq:Robin}), which reads for the Green's function as
\begin{eqnarray} \label{eq:Ga_Robin}
\bigl(\LLambda_a G_q'(r,r_0) - G_q(r,r_0)\bigr)_{|r=a} &=& 0 ,\\  
\label{eq:Gb_Robin}
\bigl(\LLambda_b G_q'(r,r_0) + G_q(r,r_0)\bigr)_{|r=b} &=& 0 .
\end{eqnarray}
If $\I(r)$ and $\K(r)$ denote two independent solutions of the
differential equation $\L u = u$, then the functions $v_q^{a,b}$ can
be expressed as their linear combinations:
\begin{eqnarray}
v_q^a(r) &=& w_q^{11} \I(qr) - w_q^{12} \K(qr) ,  \\
v_q^b(r) &=& w_q^{21} \I(qr) - w_q^{22} \K(qr) ,
\end{eqnarray}
with coefficients
\begin{eqnarray}
w_q^{11} &=& q \LLambda_a \K'(qa) - \K(qa), \quad w_q^{12} = q \LLambda_a \I'(qa) - \I(qa) , \\
w_q^{21} &=& q \LLambda_b \K'(qb) + \K(qb), \quad w_q^{22} = q \LLambda_b \I'(qb) + \I(qb) .
\end{eqnarray}
In particular, the denominator in the right-hand side of
Eq. (\ref{eq:general0}) becomes
\begin{equation}  \label{eq:wron}
v_q^b(r_0) v_q^{a\,\prime}(r_0) - v_q^a(r_0) v_q^{b\,\prime}(r_0) = - q V(q) \W(qr_0) ,
\end{equation}
where
\begin{equation}
V(q) = w_q^{11} w_q^{22} - w_q^{12} w_q^{21} 
\end{equation}
and
\begin{equation}
\W(z) = \K(z) \I'(z) - \I(z) \K'(z)
\end{equation}
is the Wronskian of two solutions.

The Robin boundary condition for an eigenfunction
$-iv_{i\sqrt{\lambda}}^a(r)$ at $r = b$ yields an equation on the
eigenvalues
\begin{equation}
V(iq) = 0 .
\end{equation}
Denoting the solutions of this equation as $q_k$ (with $k =
1,2,\ldots$), one gets the eigenvalues and eigenfunctions
\begin{equation}
\lambda_k = q_k^2 , \qquad 
u_k(r) = w_{iq_k}^{11} \, \I(iq_k r) - w_{iq_k}^{12} \,\K(iq_k r) .
\end{equation}
The general relation (\ref{eq:general0}) can thus be written as
\begin{equation}  \label{eq:general}
\sum\limits_{k=1}^\infty c_k^2 \frac{u_k(r) \, u_k^*(r_0)}{q^2 + \lambda_k} = -
\frac{v_q^b(r_0) \, v_q^a(r)}{q V(q) \W(qr_0) \omega(r_0)} \qquad (a \leq r \leq r_0 \leq b)
\end{equation}
(the other case $r \geq r_0$ is obtained in a similar way).  As the
operator $\L$ with Robin boundary conditions is self-adjoint, its
eigenfunctions $u_k(r)$ form a complete basis in the space of
square-integrable functions on $(a,b)$ with the weighting function
$\omega(r_0)$.  The completeness relation can be formally written as
\begin{equation}  \label{eq:completeness}
\omega(r_0) \sum\limits_{k=1}^\infty c_k^2 \, u_k(r) \, u_k^*(r_0) = \delta(r - r_0)  \qquad (a \leq r \leq b,~ a\leq r_0 \leq b).
\end{equation}

\subsection{One-dimensional case}
\label{sec:1D_general}

Due to the translational invariance in one dimension, diffusion on the
interval $(a,b)$ is fully equivalent to that on the shifted interval
$(0,b-a)$.  Without loss of generality, we set then $a = 0$ and
consider diffusion on $(0,b)$, which is described by the second-order
derivative, $\L_1 = \partial_r^2$.  Two independent solutions of the
equation $u'' - u = 0$ read
\begin{equation}
\I(r) = \sinh(r) , \qquad \K(r) = \cosh(r),
\end{equation}
so that
\begin{eqnarray}
w_q^{11} &=& - 1 , \quad w_q^{12} = q \LLambda_0  , \\
w_q^{21} &=& q\LLambda_b \sinh(qb) + \cosh(qb), \quad w_q^{22} = q\LLambda_b \cosh(qb) + \sinh(qb) ,
\end{eqnarray}
and
\begin{equation}
V(q) = -(q^2 \LLambda_0 \LLambda_b + 1)\sinh(qb) - q(\LLambda_0 + \LLambda_b)\cosh(qb),
\end{equation}
from which
\begin{eqnarray}  \label{eq:vqa_1D}
v_q^a(r) &=& - \sinh(qr) - q\LLambda_0 \cosh (qr), \\
v_q^b(r) &=& - \sinh(q(b-r)) - q\LLambda_b \cosh(q(b-r)) .
\end{eqnarray}
The eigenvalues are determined as $\lambda_k = q_k^2$, with $q_k$
being the positive solutions of the equation
\begin{equation}  \label{eq:alpha_1D}
V(iq) = i\bigl[(q^2\LLambda_0\LLambda_b - 1)\sin(qb) - q(\LLambda_0 + \LLambda_b)\cos(qb)\bigr] = 0,
\end{equation}
while the eigenfunctions from Eq. (\ref{eq:general_eigen}) have a form
\begin{equation}  \label{eq:1D_uk}
u_k(r) = \sin(q_k r) + q_k \LLambda_0 \cos (q_k r) .
\end{equation}
Integrating $u_k^2$ with the weighting function $\omega(r_0) = 1$ and
using trigonometric identities, one finds
\begin{eqnarray}  \label{eq:ck_1D}
c_k^2 &=& \frac{2h_0^2(\alpha_k^4 + \alpha_k^2(h_0^2+h_b^2) + h_0^2h_b^2)}{b} \biggl\{\alpha_k^6 + \alpha_k^4(2h_0^2+h_b^2+h_0+h_b) \\ \nonumber
&+& \alpha_k^2(h_0^2(h_0^2+2h_b^2)+h_0(h_b+h_0)^2) + (h_0+h_b+h_0h_b)h_bh_0^3  \biggr\}^{-1} \,,
\end{eqnarray} 
where 
\begin{equation}
\alpha_k = q_k/b,  \qquad h_0 = b/\LLambda_0, \qquad h_b = b/\LLambda_b
\end{equation}
are dimensionless quantities.  It is also convenient to rewrite
Eq. (\ref{eq:alpha_1D}) as an equation on $\alpha_k$:
\begin{equation}
\ctan \, \alpha - \frac{\alpha^2 - h_0 h_b}{\alpha (h_0 + h_b)} = 0.
\end{equation}
As the function in the left-hand side is continuous and monotonously
decreasing on an interval $(\pi (k-1),\pi k)$ from $+\infty$ to
$-\infty$ (that can be checked by computing its derivative), there is
a single zero $\alpha_k$ on each such interval, implying $\pi(k-1)
\leq \alpha_k \leq \pi k$ for $k = 1,2,3,\ldots$.

Using Eq. (\ref{eq:wron}) and $\W(z) = 1$, the summation formula
(\ref{eq:general}) becomes 
\begin{equation}  \label{eq:sum1D}
\fl
\sum\limits_{k=1}^\infty c_k^2 \frac{u_k(r) \, u_k(r_0)}{q^2 + \lambda_k} =
\frac{\bigl[\sinh(qr) + q\LLambda_0 \cosh (qr)\bigr]  \bigl[\sinh(q(b-r_0)) + q\LLambda_b \cosh(q(b-r_0))\bigr]}
{q \bigl[(q^2\LLambda_0 \LLambda_b + 1)\sinh(qb) + q(\LLambda_0 + \LLambda_b)\cosh(qb)\bigr] } 
\end{equation}  
for $0\leq r\leq r_0\leq b$.  Note that the left-hand side is
precisely the Laplace-transformed heat kernel:
\begin{equation}
\tilde{P}(r,p|r_0) = \frac{1}{D} \sum\limits_{k=1}^\infty c_k^2 \frac{u_k(r) \, u_k(r_0)}{p/D + \lambda_k} \,,
\end{equation}
so that Eq. (\ref{eq:sum1D}) yields its analytical form, with $q =
\sqrt{p/D}$.
From the identity (\ref{eq:sum1D}), one can derive various summation
formulas.  For instance, setting $r_0 = r$ and integrating over $r$
from $0$ to $b$, one finds
\begin{equation}  \label{eq:sum1D_particular}
\fl
\sum\limits_{k=1}^\infty \frac{1}{q^2 + \lambda_k} 
 = \frac{qb(q^2\LLambda_0\LLambda_b + 1)\cosh(qb) + (q^2(\LLambda_0\LLambda_b + b(\LLambda_0 + \LLambda_b))- 1)\sinh(qb)}
{2q^2 \bigl[(q^2\LLambda_0 \LLambda_b + 1)\sinh(qb) + q(\LLambda_0 + \LLambda_b)\cosh(qb)\bigr] } \,,
\end{equation}  
which is proportional to the Laplace-transformed heat trace
$\tilde{Z}(p)$.

The particular cases of the general identity (\ref{eq:sum1D})
corresponding to nine combinations of boundary conditions at two
endpoints are summarized in Table \ref{tab:summary_1D}.  For instance,
in the Neumann-Neumann case ($\LLambda_0 = \LLambda_b = \infty$), for
which $\alpha_k = \pi (k-1)$ ($k = 1,2,\ldots$), one has
\begin{equation}  \label{eq:sum1D_NN}
\fl
\sum\limits_{k=1}^\infty  \frac{\cos(\pi k x) \, \cos(\pi k x_0)}{z^2 + \pi^2 k^2} = 
\frac{\cosh(z(1-x_0)) \, \cosh(z x)}{2z \sinh z} - \frac{1}{2z^2} \qquad (0 \leq x \leq x_0 \leq 1),
\end{equation}
where we fixed $b = 1$.  Setting $x = x_0 = 0$, one retrieves
Eq. (\ref{eq:example_ctan2b}).  In the particular cases of
Dirichlet-Dirichlet and Neumann-Neumann conditions, many summation
formulas can be deduced from standard Fourier series, e.g.,
\begin{equation}
\sum\limits_{k=1}^\infty \frac{\cos (\pi k x)}{\pi k} = - \frac{1}{2\pi} \ln(2 - 2\cos(\pi x)) \qquad (0 < x \leq 1)\,.
\end{equation}

\begin{table}
\centering
\begin{turn}{90}
\begin{tabular}{|c|c|c|c|}  \hline
                           & Robin ($h_1 = h$) &   Neumann ($h_1 = 0$) &  Dirichlet ($h_1 = \infty$) \\  \hline
\multirow{4}{20mm}{\centering Robin \\ $(h_0 = h)$}
& $\tan\alpha_k = \frac{2h\alpha_k}{\alpha_k^2 - h^2}$  
& $\alpha_k \tan \alpha_k = h$  
& $\tan \alpha_k = - \alpha_k/h$ \\  
%
%
\rule{0pt}{4ex} 
& $\sum\limits_{k=1}^\infty \frac{u_k(x) \, u_k(x_0)}{(z^2 + \alpha_k^2)(\alpha_k^2 + 2h + h^2)}$  
& $\sum\limits_{k=1}^\infty \frac{u_k(x) \, u_k(x_0)}{(z^2 + \alpha_k^2) (\alpha_k^2 + h + h^2)}$ 
& $\sum\limits_{k=1}^\infty  \frac{u_k(x) \, u_k(x_0)}{(z^2 + \alpha_k^2)(\alpha_k^2 + h + h^2)}$ \\  
& $= \frac{v(1-x_0; z,h) \, v(x; z,h)} {2z \bigl[(z^2+h^2) \sinh z + 2hz \cosh z\bigr]}$  
& $= \frac{v(x; z,h) \, \cosh(z(1-x_0))}{2z \bigl[z \sinh z + h \cosh z\bigr]}$   
& $= \frac{\sinh(z(1-x_0)) \, v(x; z,h)}{2z \bigl[h \sinh z + z \cosh z\bigr]}$ \\   
\rule{0pt}{4ex} 
& $Z = \frac{z(z^2+h^2)\cosh z + (z^2(1+2h)-h^2) \sinh z}{2z^2 ((z^2+h^2)\sinh z + 2zh \cosh z)}$ 
& $Z = \frac{z\cosh z + (h+1) \sinh z}{2z(z\sinh z + h \cosh z)}$  
& $Z = \frac{hz\cosh z + (z^2 - h) \sinh z}{2z^2(h\sinh z + z\cosh z)}$ \\ \hline   
%
\multirow{4}{20mm}{\centering Neumann \\ $(h_0 = 0)$}
& $\alpha_k \tan \alpha_k = h$   
& $\alpha_k = \pi(k-1)$   
& $\alpha_k = \pi(k-1/2)$ \\ 
%
%
\rule{0pt}{4ex} 
& $\sum\limits_{k=1}^\infty \frac{\cos(\alpha_k x) \, \cos(\alpha_k x_0) \, (\alpha_k^2 + h^2)}{(z^2 + \alpha_k^2)(\alpha_k^2 + h + h^2)}$ 
& $\frac{1}{2z^2} + \sum\limits_{k=2}^\infty \frac{\cos(\alpha_k x) \, \cos(\alpha_k x_0)}{z^2 + \alpha_k^2}$  
& $\sum\limits_{k=1}^\infty \frac{\cos(\alpha_k x) \, \cos(\alpha_k x_0)}{z^2 + \alpha_k^2}$\\ 
& $= \frac{v(1-x_0;z;h) \,  \cosh(z x)}{2z \bigl[ z \sinh z + h \cosh z\bigr]}$ 
& $= \frac{\cosh(z(1-x_0)) \,  \cosh(z x)}{2z \sinh z}$  
& $= \frac{\sinh(z(1-x_0)) \,  \cosh(z x)}{2z \cosh z}$\\   
\rule{0pt}{4ex} 
& $Z = \frac{z\cosh z + (h+1) \sinh z}{2z(z\sinh z + h \cosh z)}$  
& $Z = \frac{z\cosh z + \sinh z}{2z^2 \sinh z}$   
& $Z = \frac{\sinh z}{2z \cosh z}$ \\   \hline  
\multirow{4}{20mm}{\centering Dirichlet \\ $(h_0 = \infty)$}
& $\tan \alpha_k = - \alpha_k/h$   
& $\alpha_k = \pi(k-1/2)$  
& $\alpha_k = \pi k$ \\   
%
%
\rule{0pt}{4ex} 
& $\sum\limits_{k=1}^\infty  \frac{\sin(\alpha_k x) \, \sin(\alpha_k x_0) \, (\alpha_k^2 + h^2)}{(z^2 + \alpha_k^2)(\alpha_k^2 + h + h^2)}$  
& $\sum\limits_{k=1}^\infty  \frac{\sin(\alpha_k x) \, \sin(\alpha_k x_0)}{z^2 + \alpha_k^2}$  
& $\sum\limits_{k=1}^\infty  \frac{\sin(\alpha_k x) \, \sin(\alpha_k x_0)}{z^2 + \alpha_k^2}$ \\  
& $= \frac{v(1-x_0; z,h) \, \sinh(z x)} {2z \bigl[h \sinh z + z \cosh z\bigr]}$ 
& $= \frac{\cosh(z(1-x_0)) \, \sinh(z x)}{2z \cosh z}$  
& $= \frac{\sinh(z(1-x_0)) \, \sinh(z x)}{2z \sinh z}$ \\  
\rule{0pt}{4ex} 
& $Z = \frac{hz\cosh z + (z^2 - h) \sinh z}{2z^2(h\sinh z + z\cosh z)}$  
& $Z = \frac{\sinh z}{2z\cosh z}$   
& $Z = \frac{\cosh z}{2z \sinh z} - \frac{1}{2z^2}$ \\ \hline   
\end{tabular}
\end{turn}
\caption{
Summary of summation formulas for nine combinations of boundary
conditions at the endpoints of the unit interval ($b = 1$).  Here $k =
1,2,\ldots$, $0 \leq x \leq x_0 \leq 1$, $0 < h < \infty$, $z$ is a
complex number, and $Z = \sum\nolimits_{k=1}^\infty \frac{1}{z^2 +
\alpha_k^2}$.  For convenience, we rescale Eqs. (\ref{eq:1D_uk},
\ref{eq:vqa_1D}) by $h$ to get: $u_k(x) = h \sin(\alpha_k x) +
\alpha_k \cos(\alpha_k x)$ and $v(x;z,h) = h \sinh(zx) + z\cosh(zx)$.}
\label{tab:summary_1D}
\end{table}

\subsection{Two-dimensional case}

In two dimensions, the problem $\L_2 u - u = 0$ with the radial
operator $\L_2$ from Eq. (\ref{eq:L_radial}) is reduced to the
modified Bessel equation whose two independent solutions are the
modified Bessel functions $I_n(z)$ and $K_n(z)$.  While most of the
following results are applicable for $n \in \R$ (or even $n \in \C$),
we restrict our discussion to the case of an integer $n$, which is
most relevant for the Laplace operator in two-dimensional
rotation-invariant domains.

\subsubsection{Circular annulus}

In the following, the general results are applied for a fixed $n$ by
setting
\begin{equation}
\I(r) = I_n(r), \qquad \K(r) = K_n(r),
\end{equation}
Using the relations
\begin{equation}
I_n(-iz) = (-i)^n J_n(z), \qquad  K_n(-iz) = \frac{\pi i^{n+1}}{2} (J_n(z) + i Y_n(z)),
\end{equation}
one gets
\begin{equation}
\fl
v_{-iq}^b(r) = - \frac{\pi}{2} \biggl( \bigl(qY'_n(qb) + h_b Y_n(qb) \bigr) J_n(qr) - \bigl(qJ'_n(qb) + h_b J_n(qb) \bigr) Y_n(qr) \biggr).
\end{equation}
For convenience, we change a numerical factor in our former expression
for eigenfunctions to write
\begin{eqnarray}
u_{nk}(r) &=& \bigl(q_{nk} \LLambda_b Y'_n(q_{nk} b) + Y_n(q_{nk} b) \bigr) J_n(q_{nk} r)  \\  \nonumber
&-& \bigl(q_{nk} \LLambda_b J'_n(q_{nk} b) + J_n(q_{nk} b) \bigr) Y_n(q_{nk} r),
\end{eqnarray}
where $q_{nk}$ are the positive solutions of the equation
\begin{eqnarray} \nonumber
&& \bigl(q \LLambda_a Y'_n(qa) - Y_n(qa)\bigr) \bigl(q \LLambda_b J'_n(qb) + J_n(qb)\bigr) \\
&& - \bigl(q \LLambda_a J'_n(qa) - J_n(qa)\bigr) (q \LLambda_b Y'_n(qb) + Y_n(qb)\bigr) = 0,
\end{eqnarray}
enumerated by the index $k = 1,2,3,\ldots$.  The Laplacian eigenvalues
are $\lambda_{nk} = q_{nk}^2$.  The standard computation yields the
$L_2$-normalization coefficients of the eigenfunction with the
weighting function $\omega(r_0) = r_0$
\begin{eqnarray}
c_{nk}^2 &=& 2q_{nk}^2 \left\{ \biggl( b^2 (u'_{nk}(b))^2 + (q_{nk}^2b^2 - n^2) (u_{nk}(b))^2 \biggr) \right. \\  \nonumber
&-& \left. \biggl( a^2 (u'_{nk}(a))^2 + (q_{nk}^2 a^2 - n^2) (u_{nk}(a))^2 \biggr) \right\}^{-1} .
\end{eqnarray}    

Using the Wronskian of the functions $I_n(z)$ and $K_n(z)$, 
\begin{equation}
\W(z) = I'_n(z) K_n(z) - I_n(z) K'_n(z) = \frac{1}{z} \,, 
\end{equation}
one rewrites Eq. (\ref{eq:general}) for each $n$ as
\begin{equation}  \label{eq:general_2D}
\sum\limits_{k=1}^\infty c_{nk}^2 \frac{u_{nk}(r) \, u_{nk}^*(r_0)}{q^2 + \lambda_{nk}} = -
\frac{v_{n,q}^b(r_0) \, v_{n,q}^a(r)}{V_n(q)}  \qquad (a \leq r \leq r_0 \leq b)\,,
\end{equation}    
where
\begin{eqnarray}
\fl
v_{n,q}^a(r) &=& \bigl(q\LLambda_a K'_n(qa) - K_n(qa)\bigr) \, I_n(qr) - \bigl(q\LLambda_a I'_n(qa) - I_n(qa)\bigr) \, K_n(qr), \\
\fl
v_{n,q}^b(r) &=& \bigl(q\LLambda_b K'_n(qb) + K_n(qb)\bigr) \, I_n(qr) - \bigl(q\LLambda_b I'_n(qb) + I_n(qb)\bigr) \, K_n(qr),
\end{eqnarray}
and
\begin{eqnarray}
V_n(q) &=& \bigl(q\LLambda_a K'_n(qa) - K_n(qa)\bigr) \bigl(q\LLambda_b I'_n(qb) + I_n(qb)\bigr) \\  \nonumber
&-& \bigl(q\LLambda_b K'_n(qb) + K_n(qb)\bigr) \bigl(q\LLambda_a I'_n(qa) - I_n(qa)\bigr) .
\end{eqnarray}
Eq. (\ref{eq:general_2D}) is the main summation formula for circular
annuli.

Setting $r_0 = r$ and integrating over $r$ from $a$ to $b$ with the
weighting function $\omega(r_0) = r_0$, one deduces
\begin{equation}  \label{eq:general_2D_special0}
\sum\limits_{k=1}^\infty \frac{1}{q^2 + \lambda_{nk}} = \frac{S_n(a) - S_n(b)}{2q^2 V_n(q)}  \,,
\end{equation}    
with
\begin{equation}
S_n(r) = (q^2 r^2 + n^2)v_{n,q}^a(r) v_{n,q}^b(r) - r^2 v_{n,q}^{a\, \prime}(r) v_{n,q}^{b\, \prime}(r) ,
\end{equation}
where we employed the modified Bessel equation for functions $I_n(z)$
and $K_n(z)$ to calculate the integral:
\begin{equation}
\int\limits_a^b dr \, r \, v_{n,q}^a(r) \, v_{n,q}^b(r) = \frac{S_n(b) - S_n(a)}{2q^2} \,.
\end{equation}

Using the explicit form of the $L_2$-normalized Laplacian
eigenfunctions in polar coordinates,
\begin{equation}
U_{nk}(r,\phi) = \frac{1}{\sqrt{2\pi}} \, c_{nk}\, u_{nk}(r) \, e^{in\phi} ,
\end{equation}
one rewrites the spectral representations of the heat kernel and
Laplace-transformed heat kernel in $\Omega$ as
\begin{equation}  \label{eq:Pt_2D}
P(\x,t|\x_0) = \frac{1}{2\pi} \sum\limits_{n=-\infty}^\infty e^{in(\phi-\phi_0)} 
\sum\limits_{k=1}^\infty c_{nk}^2 \, u_{nk}(r) \, u_{nk}(r_0) \, e^{-Dt\lambda_{nk}} 
\end{equation}
and
\begin{equation}  \label{eq:Gq_2D}
\tilde{P}(\x,p|\x_0) = \frac{1}{2\pi} \sum\limits_{n=-\infty}^\infty e^{in(\phi-\phi_0)} 
\sum\limits_{k=1}^\infty c_{nk}^2 \frac{u_{nk}(r) \, u_{nk}(r_0)}{p + D\lambda_{nk}} \,. 
\end{equation}
On the other hand, using the summation formula (\ref{eq:general_2D}),
one gets an alternative, more explicit representation for $a \leq r
\leq r_0 \leq b$:
\begin{equation}
\tilde{P}(\x,p|\x_0) = - \frac{1}{2\pi D} \sum\limits_{n=-\infty}^\infty e^{in(\phi - \phi_0)}
\frac{v_{n,q}^b(r_0) \, v_{n,q}^a(r)}{V_n(q)} \,,
\end{equation}
with $q = \sqrt{p/D}$.

\subsubsection{Disk} 

For the disk of radius $b$, one takes the limit $a\to 0$, for which
$w_q^{11} = q\LLambda_a K'_n(qa) - K_n(qa)$ diverges that simplifies
various expressions.  Table \ref{tab:DS} summarizes most relevant
summation formulas for both Dirichlet and Robin/Neumann boundary
conditions.  In particular, Eqs. (\ref{eq:general_2D}) and
(\ref{eq:general_2D_special0}) are reduced to Eqs. (D1,D2) from that
table.
%
At $x_0 = 1$, Eq. (D1) yields Eq. (D3).
Setting $x = 1$, substituting $z$ by $\sqrt{s}$ and differentiating by
$s$ this identity $m$ times, one recovers Eq. (\ref{eq:Sneddon_S})
after an explicit evaluation of the derivatives in the right-hand
side.  The identities (D4-D6) are obtained by replacing $z$ by $-iz$
in Eqs. (D1-D3), whereas the identities (D7-D12) are deduced from
Eqs. (D1-D6) in the limit $\LLambda_b\to 0$ or $h = b/\LLambda_b
\to\infty$ (Dirichlet condition).
Note that Eqs. (D9) and (D12) can alternatively be written as
\begin{equation}  \label{eq:general_2D0_specD2}
\frac{I_n(zx)}{I_n(z)} = x^n + 2z^2 \sum\limits_{k=1}^\infty \frac{J_n(\alpha_{nk}x)}{(z^2 + \alpha_{nk}^2) \alpha_{nk} J'_n(\alpha_{nk})} \,.
\end{equation}
and
%
\begin{equation}  \label{eq:general_2D0_specDi}
\frac{J_n(zx)}{J_n(z)} = x^n + 2z^2 \sum\limits_{k=1}^\infty \frac{J_n(\alpha_{nk}x)}{(z^2 - \alpha_{nk}^2) \alpha_{nk} J'_n(\alpha_{nk})} \,.
\end{equation}
Note also that Eq. (D8) reproduces Eq. (\ref{eq:eta_1}) for the disk,
whereas Eq. (D10) is a particular form of the Kneser-Sommerfeld
expansion.%
\footnote{
As indicated in \cite{Hayashi81}, the Kneser-Sommerfeld expansion
provided by Watson (see \cite{Watson}, p. 499) has to be corrected.
We checked numerically that Watson's expression is indeed not valid.}

We also note that the completeness relation (\ref{eq:completeness}) can be
written explicitly in the case of the unit disk ($b = 1$) as
\begin{equation}
\delta(x - x_0) = 2x_0 \sum\limits_{k=1}^\infty \frac{J_n(\alpha_{nk} x)\, J_n(\alpha_{nk} x_0) \, \alpha_{nk}^2}
{J_n^2(\alpha_{nk}) (\alpha_{nk}^2 - n^2 + h^2)}  \qquad 
\left( \begin{array}{c} 0 \leq x \leq 1 \\ 0 \leq x_0 \leq 1 \\ \end{array} \right).  
\end{equation}
Multiplying this relation by a function $f(x_0)$ and integrating over
$x_0$ from $0$ to $1$, one retrieves the Dini expansion
(\ref{eq:Dini}).  Similarly, one retrieves the Fourier-Bessel
expansions (\ref{eq:FB_Dirichlet}, \ref{eq:FB_Neumann}) in the
Dirichlet ($h = \infty$) and Neumann ($h = 0$) cases.  As previously,
we skip the mathematical constraints needed to ensure the convergence
of these expansions (see Sec. \ref{sec:Dini} and
\cite{Watson}).

\begin{table}   
\centering
\begin{turn}{90}
\begin{tabular}{|c|lr|lr|}  \hline
 & \hskip 40mm Disk && \hskip 45mm  Ball &\\  \hline
\multirow{8}{6mm}[-12mm]{\begin{turn}{270} Robin ($0\leq h<\infty$) \end{turn}}
& $\sum\limits_{k=1}^\infty \frac{2\alpha_{nk}^2 \, J_n(\alpha_{nk} x) \, J_n(\alpha_{nk} x_0)}
{(\alpha_{nk}^2 - n^2 + h^2)(z^2 + \alpha_{nk}^2)\, J_n^2(\alpha_{nk})} = $ & (D1)
& $\sum\limits_{k=1}^\infty \frac{2\alpha_{nk}^2 \, j_n(\alpha_{nk} x) \, j_n(\alpha_{nk} x_0)}
{(\alpha_{nk}^2 - n(n+1) + h^2 - h)(z^2 + \alpha_{nk}^2) \, j_n^2(\alpha_{nk})} =$ & (S1)  \\ 
& $\biggl(K_n(zx_0) - I_n(zx_0) \frac{z K'_n(z) + h K_n(z)}{z I'_n(z) + h I_n(z)} \biggr) I_n(zx)$ &
& $z \biggl(k_n(zx_0) - i_n(zx_0) \frac{z k'_n(z) + h k_n(z)}{z i'_n(z) + h i_n(z)} \biggr) i_n(zx)$ & \\    
& $\sum\limits_{k=1}^\infty \frac{1}{z^2 + \alpha_{nk}^2} = 
\frac{(z^2 + n^2) I_n(z) + z h I'_n(z)}{2z^2(z I'_n(z) + h I_n(z))} - \frac{n}{2z^2}$ & (D2)
& $\sum\limits_{k=1}^\infty \frac{1}{z^2 + \alpha_{nk}^2} = 
\frac{(z^2 + n(n+1)) i_n(z) + z(h-1) i'_n(z)}{2z^2(z i'_n(z) + h i_n(z))} - \frac{n}{2z^2}$ & (S2) \\ 
%
& $\sum\limits_{k=1}^\infty \frac{2\alpha_{nk}^2 \, J_n(\alpha_{nk}x)}{(\alpha_{nk}^2 - n^2 + h^2) (z^2 + \alpha_{nk}^2) J_n(\alpha_{nk})} 
= \frac{I_n(z{\clr x})}{z I'_n(z) + h I_n(z)}$ & (D3) 
& $\sum\limits_{k=1}^\infty \frac{2\alpha_{nk}^2 \, j_n(\alpha_{nk}x)}{(\alpha_{nk}^2 - n(n+1) + h^2-h)(z^2 + \alpha_{nk}^2) \, j_n(\alpha_{nk})} 
= \frac{i_n(z{\clr x})}{z i'_n(z) + h i_n(z)}$ & (S3) \\  \cline{2-5}
& $\sum\limits_{k=1}^\infty \frac{2\alpha_{nk}^2 \, J_n(\alpha_{nk} x) \, J_n(\alpha_{nk} x_0)}
{(\alpha_{nk}^2 - n^2 + h^2)(z^2 - \alpha_{nk}^2)\, J_n^2(\alpha_{nk})} = $ & (D4) 
& $\sum\limits_{k=1}^\infty \frac{2\alpha_{nk}^2 \, j_n(\alpha_{nk} x) \, j_n(\alpha_{nk} x_0)}
{(\alpha_{nk}^2 - n(n+1) + h^2 - h)(z^2 - \alpha_{nk}^2) \, j_n^2(\alpha_{nk})} =$ & (S4) \\ 
& ${\clr \frac{\pi}{2}}\biggl(Y_n(zx_0) - J_n(zx_0) \frac{z Y'_n(z) + h Y_n(z)}{z J'_n(z) + h J_n(z)} \biggr) J_n(zx)$ &
& $z \biggl(y_n(zx_0) - j_n(zx_0) \frac{z y'_n(z) + h y_n(z)}{z j'_n(z) + h j_n(z)} \biggr) j_n(zx)$ & \\   
& $\sum\limits_{k=1}^\infty \frac{1}{z^2 - \alpha_{nk}^2} = 
\frac{(n^2 - z^2) J_n(z) + z h J'_n(z)}{2z^2(z J'_n(z) + h J_n(z))} - \frac{n}{2z^2}$ & (D5)
& $\sum\limits_{k=1}^\infty \frac{1}{z^2 - \alpha_{nk}^2} = 
\frac{(n(n+1)- z^2) j_n(z) + z(h-1) j'_n(z)}{2z^2(z j'_n(z) + h j_n(z))} - \frac{n}{2z^2}$ & (S5) \\  
& $\sum\limits_{k=1}^\infty \frac{2\alpha_{nk}^2 \, J_n(\alpha_{nk}x)}{(\alpha_{nk}^2 - n^2 + h^2) (z^2 - \alpha_{nk}^2) J_n(\alpha_{nk})} 
= \frac{-J_n(z{\clr x})}{z J'_n(z) + h J_n(z)}$ & (D6)
& $\sum\limits_{k=1}^\infty \frac{2\alpha_{nk}^2 \, j_n(\alpha_{nk}x)}{(\alpha_{nk}^2 - n(n+1) + h^2-h)(z^2 - \alpha_{nk}^2) \, j_n(\alpha_{nk})} 
= \frac{-j_n(z{\clr x})}{z j'_n(z) + h j_n(z)}$ & (S6) \\  \hline \hline
\multirow{8}{6mm}[-10mm]{\begin{turn}{270} Dirichlet ($h = \infty$) \end{turn}}
& $\sum\limits_{k=1}^\infty \frac{2 J_n(\alpha_{nk} x) \, J_n(\alpha_{nk} x_0)}
{(z^2 + \alpha_{nk}^2)\, [J'_n(\alpha_{nk})]^2} = $ & (D7)
& $\sum\limits_{k=1}^\infty \frac{2 j_n(\alpha_{nk} x) \, j_n(\alpha_{nk} x_0)}
{(z^2 + \alpha_{nk}^2) \, [j'_n(\alpha_{nk})]^2} =$ & (S7) \\ 
& $\biggl(K_n(zx_0) - I_n(zx_0) \frac{K_n(z)}{I_n(z)} \biggr) I_n(zx)$ &
& $z \biggl(k_n(zx_0) - i_n(zx_0) \frac{k_n(z)}{i_n(z)} \biggr) i_n(zx)$ & \\  
& $\sum\limits_{k=1}^\infty \frac{1}{z^2 + \alpha_{nk}^2} = 
\frac{I'_n(z)}{2z I_n(z)} - \frac{n}{2z^2}$ & (D8)
& $\sum\limits_{k=1}^\infty \frac{1}{z^2 + \alpha_{nk}^2} = 
\frac{i'_n(z)}{2z i_n(z)} - \frac{n}{2z^2}$ & (S8) \\ 
%
& $\sum\limits_{k=1}^\infty \frac{\alpha_{nk} \, J_n(\alpha_{nk}x)}{(z^2 + \alpha_{nk}^2) \,J'_n(\alpha_{nk})} 
= -\frac{I_n(zx)}{2I_n(z)}$ & (D9)
& $\sum\limits_{k=1}^\infty \frac{\alpha_{nk} \, j_n(\alpha_{nk}x)}{(z^2 + \alpha_{nk}^2) \,j'_n(\alpha_{nk})} 
= -\frac{i_n(zx)}{2i_n(z)}$ & (S9) \\  \cline{2-5}
& $\sum\limits_{k=1}^\infty \frac{2 J_n(\alpha_{nk} x) \, J_n(\alpha_{nk} x_0)}
{(z^2 - \alpha_{nk}^2)\, [J'_n(\alpha_{nk})]^2} = $ & (D10)
& $\sum\limits_{k=1}^\infty \frac{2 j_n(\alpha_{nk} x) \, j_n(\alpha_{nk} x_0)}
{(z^2 - \alpha_{nk}^2) \, [j'_n(\alpha_{nk})]^2} =$ & (S10) \\ 
& ${\clr \frac{\pi}{2}}\biggl(Y_n(zx_0) - J_n(zx_0) \frac{Y_n(z)}{J_n(z)} \biggr) J_n(zx)$ & 
& $z \biggl(y_n(zx_0) - j_n(zx_0) \frac{y_n(z)}{j_n(z)} \biggr) j_n(zx)$ &  \\  
& $\sum\limits_{k=1}^\infty \frac{1}{z^2 - \alpha_{nk}^2} = \frac{J'_n(z)}{2z J_n(z)} - \frac{n}{2z^2}$ & (D11)
& $\sum\limits_{k=1}^\infty \frac{1}{z^2 - \alpha_{nk}^2} = \frac{j'_n(z)}{2z j_n(z)} - \frac{n}{2z^2}$ & (S11) \\  
& $\sum\limits_{k=1}^\infty \frac{\alpha_{nk} \, J_n(\alpha_{nk}x)}{(z^2 - \alpha_{nk}^2) \,J'_n(\alpha_{nk})} 
= {\clr +}\frac{J_n(zx)}{2J_n(z)}$ & (D12)
& $\sum\limits_{k=1}^\infty \frac{\alpha_{nk} \, j_n(\alpha_{nk}x)}{(z^2 - \alpha_{nk}^2) \,j'_n(\alpha_{nk})} 
= {\clr +}\frac{j_n(zx)}{2j_n(z)}$ & (S12) \\  \hline
\end{tabular}
\end{turn}
\caption{
Summary of summation formulas for a disk and a ball of radius $b$.
Here $0 \leq x \leq x_0 \leq 1$, and $\alpha_{nk} = bq_{nk}$ are the
positive zeros of $g_n(z) = zJ'_n(z) + hJ_n(z)$ in two dimensions, and
of $g_n(z) = zj'_n(z) + hj_n(z)$ in three dimensions, with $h =
b/\LLambda_b$.  Expressions for Neumann boundary condition are simply
deduced by setting $h= 0$. {\it Several misprints were found in the
published version of this Table; they are corrected here and highligted
by red color.}}
\label{tab:DS}
\end{table}


\subsection{Three-dimensional case}

In three dimensions, a substitution $u(r) = \bar{u}(r)/\sqrt{r}$
reduces the eigenvalue equation $\L_3 u = u$ with the radial operator
$\L_3$ from Eq. (\ref{eq:L_radial}) to the modified Bessel equation
\begin{equation}
r^2 \bar{u}'' + r \bar{u}' - [r^2 + (n+1/2)^2] \bar{u} = 0 .
\end{equation}
As a consequence, one can formally extend most of the results of the
previous subsection by substituting $u(r)$ by $\sqrt{r} u(r)$ and $n$
by $n+1/2$.  

\subsubsection{Spherical shell}

Two linearly independent solutions of the equation $\L_3 u = u$ are
the modified spherical Bessel functions, $i_n(z)$ and $k_n(z)$, which
are related to the modified Bessel functions $I_\nu(z)$ and $K_\nu(z)$
as
\begin{equation}  \label{eq:sBessel_ik}
i_n(z) = \sqrt{\pi/(2z)} \, I_{n+1/2}(z), \qquad
k_n(z) = \sqrt{2/(\pi z)} \, K_{n+1/2}(z) .
\end{equation}
For each fixed $n$, we set therefore
\begin{equation}
\I(x) = i_n(x) , \qquad  \K(x) = k_n(x).
\end{equation}

Using the following relations
\begin{equation}
i_n(iz) = i^n j_n(z),  \qquad  k_n(iz) = - (-i)^n j_n(z) - (-i)^{n+1} y_n(z),
\end{equation}
where
\begin{equation}
j_n(z) = \sqrt{\pi/(2z)} \, J_{n+1/2}(z), \qquad
y_n(z) = \sqrt{\pi/(2z)} \, Y_{n+1/2}(z) 
\end{equation}
are the spherical Bessel functions of the first and second kind, we
get, after simplifications,
\begin{eqnarray}
u_{nk}(r) &=& \bigl(q_{nk}\LLambda_a y'_n(q_{nk} a) - y_n(q_{nk} a)\bigr) \, j_n(q_{nk} r) \\  \nonumber
&-& \bigl(q_{nk}\LLambda_a j'_n(q_{nk} a) - j_n(q_{nk} a)\bigr) \, y_n(q_{nk} r),
\end{eqnarray}
where $q_{nk}$ are the positive solutions of the equation
\begin{eqnarray} \nonumber
&& \bigl(q\LLambda_a y'_n(qa) - y_n(qa)\bigr) \bigl(q\LLambda_b j'_n(qb) + j_n(qb)\bigr) \\
&& - \bigl(q\LLambda_a j'_n(qa) - j_n(qa)\bigr) (q\LLambda_b y'_n(qb) + y_n(qb)\bigr) = 0,
\end{eqnarray}
enumerated by the index $k = 1,2,3,\ldots$.  The Laplacian eigenvalues
are $\lambda_{nk} = q_{nk}^2$.

The standard computation yields the $L_2$ normalization of the
eigenfunctions with the weighting function $\omega(r_0) = r_0^2$:
\begin{eqnarray}
\fl
c_{nk}^2 &=& 2q_{nk}^2 \left\{ b \biggl[ b^2 (u'_{nk}(b))^2 + b u_{nk}(b) u'_{nk}(b) + (b^2 q_{nk}^2 - n(n+1)) u_{nk}^2(b) \biggr] \right. \\  \nonumber
\fl
&-& \left. a \biggl[ a^2 (u'_{nk}(a))^2 + a u_{nk}(a) u'_{nk}(a) + (a^2 q_{nk}^2 - n(n+1)) u_{nk}^2(a) \biggr] \right\}^{-1} .
\end{eqnarray}
Using the Wronskian of the functions $i_n(z)$ and $k_n(z)$, 
\begin{equation}
\W(z) = i'_n(z) k_n(z) - i_n(z) k'_n(z) = \frac{1}{z^2} \,, 
\end{equation}
one rewrites Eq. (\ref{eq:general}) for each $n$ as
\begin{equation}  \label{eq:general_3D}
\sum\limits_{k=1}^\infty c_{nk}^2 \frac{u_{nk}(r) \, u_{nk}^*(r_0)}{q^2 + \lambda_{nk}} = -
q \frac{v_{n,q}^b(r_0) \, v_{n,q}^a(r)}{V_n(q)} \qquad (a \leq r \leq r_0 \leq b)\,,
\end{equation}
where
\begin{eqnarray}
v_{n,q}^a(r) &=& \bigl(q\LLambda_a k'_n(qa) - k_n(qa)\bigr) \, i_n(qr) - \bigl(q\LLambda_a i'_n(qa) - i_n(qa)\bigr) \, k_n(qr), \\
v_{n,q}^b(r) &=& \bigl(q\LLambda_b k'_n(qb) + k_n(qb)\bigr) \, i_n(qr) - \bigl(q\LLambda_b i'_n(qb) + i_n(qb)\bigr) \, k_n(qr),
\end{eqnarray}
and
\begin{eqnarray}
V_n(q) &=& \bigl(q\LLambda_a k'_n(qa) - k_n(qa)\bigr) \bigl(q\LLambda_b i'_n(qb) + i_n(qb)\bigr) \\  \nonumber
&-& \bigl(q\LLambda_b k'_n(qb) + k_n(qb)\bigr) \bigl(q\LLambda_a i'_n(qa) - i_n(qa)\bigr) .
\end{eqnarray}
Eq. (\ref{eq:general_3D}) is the main summation formula for spherical
shells.

Setting $r_0 = r$ and integrating over $r$ from $a$ to $b$ with the
weighting function $\omega(r_0) = r_0^2$, one deduces
\begin{equation}  \label{eq:general_3D_special0}
\sum\limits_{k=1}^\infty \frac{1}{q^2 + \lambda_{nk}} = \frac{S_n(a) - S_n(b)}{2qV_n(q)} \,,
\end{equation}    
with
\begin{equation}
\fl
S_n(r) = r \biggl((q^2 r^2 + n(n+1))v_{n,q}^a(r) v_{n,q}^b(r) 
- \frac{r}{2} \partial_r \bigl( v_{n,q}^{a}(r) v_{n,q}^{b}(r) \bigr)
- r^2 v_{n,q}^{a\, \prime}(r) v_{n,q}^{b\, \prime}(r)\biggr),
\end{equation}
where we employed the modified Bessel equation for functions $i_n(z)$
and $k_n(z)$ to calculate the integral:
\begin{equation}
\int\limits_a^b dr \, r^2 \, v_{n,q}^a(r) \, v_{n,q}^b(r) = \frac{S_n(b) - S_n(a)}{2q^2}  \,.
\end{equation}  

Using the explicit form of the $L_2$-normalized Laplacian
eigenfunctions in spherical coordinates,
\begin{equation}
U_{nkl}(r,\theta,\phi) = \sqrt{\frac{(2n+1)(n-l)!}{4\pi (n+l)!}} \, c_{nk}\, u_{nk}(r) \, P_n^l(\cos\theta)\, e^{il\phi} ,
\end{equation}
one rewrites the spectral representation of the heat kernel and
Laplace-transformed heat kernel in $\Omega$ explicitly as:
\begin{equation}
\fl
P(\x,t|\x_0) = \frac{1}{4\pi} \sum\limits_{n=0}^\infty (2n+1) P_n\left(\frac{(\x \cdot \x_0)}{\|\x\| \, \|\x_0\|} \right)
\sum\limits_{k=1}^\infty  c_{nk}^2 \, u_{nk}(r) \, u_{nk}(r_0) \, e^{-Dt\lambda_{nk}} 
\end{equation}
and
\begin{equation}  \label{eq:Gq_3D}
\fl
\tilde{P}(\x,p|\x_0) = \frac{1}{4\pi} \sum\limits_{n=0}^\infty (2n+1) P_n\left(\frac{(\x \cdot \x_0)}{\|\x\| \, \|\x_0\|} \right)
\sum\limits_{k=1}^\infty c_{nk}^2 \, \frac{u_{nk}(r) \, u_{nk}(r_0)}{p + D\lambda_{nk}} \,,
\end{equation}
where the sum over $l$ from $-n$ to $n$ yielded, via the addition
theorem for spherical harmonics, the Legendre polynomial
$P_n\biggl(\frac{(\x \cdot \x_0)}{\|\x\| \, \|\x_0\|} \biggr)$ of the
cosine of the angle between the vectors $\x$ and $\x_0$:
\begin{equation}
\sum\limits_{l=-n}^n \frac{(n-l)!}{(n+l)!} \, P_n^l(\cos\theta) e^{il\phi}  \, P_n^l(\cos\theta_0) e^{-il\phi_0} 
= P_n\biggl(\frac{(\x \cdot \x_0)}{\|\x\|\, \|\x_0\|}\biggr) .
\end{equation}

On the other hand, using the summation formula (\ref{eq:general_3D}),
one gets an alternative, more explicit representation for $a \leq r
\leq r_0 \leq b$:
\begin{equation}
\tilde{P}(\x,p|\x_0) = - \frac{q}{4\pi D} \sum\limits_{n=0}^\infty (2n+1) P_n\left(\frac{(\x \cdot \x_0)}{\|\x\| \, \|\x_0\|}\right)
\frac{v_{n,q}^b(r_0) \, v_{n,q}^a(r)}{V_n(q)} \,,
\end{equation}
with $q = \sqrt{p/D}$.

\subsubsection{Ball}

For the ball of radius $b$, one takes the limit $a\to 0$, for which
$w_q^{11} = q\LLambda_a k'_n(qa) - k_n(qa)$ diverges that simplifies
various expressions.  Table \ref{tab:DS} summarizes the most relevant
summation formulas for both Dirichlet and Robin/Neumann boundary
conditions.  In particular, Eqs. (\ref{eq:general_3D}) and
(\ref{eq:general_3D_special0}) are reduced to Eqs. (S1,S2) from that
table.
%
At $x_0 = 1$, Eq. (S1) yields Eq. (S3).  Replacing $z$ by $-iz$ yields
Eqs. (S4-S6), while the Dirichlet limit $\LLambda_b\to 0$ or $h =
b/\LLambda_b \to\infty$ transforms Eqs. (S1-S6) to Eqs. (S7-S12).
%
Note that Eq. (S2) reproduces Eq. (\ref{eq:eta_1}) for the ball,
whereas Eq. (S4) is a Kneser-Sommerfeld expansion for spherical Bessel
functions.
%
We also note that the completeness relation (\ref{eq:completeness})
can be written explicitly in the case of the unit ball ($b = 1$) as
\begin{equation}
\fl
\delta(x - x_0) = 2x_0^2 \sum\limits_{k=1}^\infty \frac{j_n(\alpha_{nk} x)\, j_n(\alpha_{nk} x_0) \, \alpha_{nk}^2}
{j^2_n(\alpha_{nk}) (\alpha_{nk}^2 - n(n+1)+ h^2 - h)}  \qquad 
\left( \begin{array}{c} 0 \leq x \leq 1 \\ 0 \leq x_0 \leq 1 \\ \end{array} \right). 
\end{equation}

\section{Discussion and Conclusion}

Simple domains remain toy models and emblematic examples in most
applications.  While such ``spherical cows'' cannot capture the
geometric complexity of most biological, physical and engineering
systems, they bring conceptual understanding of the underlying
processes.  Moreover, simple domains often play a role of building
blocks for modeling more sophisticated environments or simulating
diffusion and other processes \cite{Torquato91,Galanti16,Grebenkov19}.
For instance, the implementation of a fast random walk algorithm
\cite{Muller56}, in which the variable size of each displacement of a
particle is adapted to the local environment, relies on the knowledge
of diffusion inside balls (see
\cite{Grebenkov14} for a short review).  This efficient algorithm was
broadly used to model diffusion-limited growth phenomena
\cite{Meakin85,Ossadnik91,Bowler05}, diffusion-reaction in porous media
\cite{Torquato89,Zheng89}, harmonic measure and other first-passage
statistics \cite{Grebenkov05,Grebenkov05b}, diffusion-weighted signal
in NMR \cite{Grebenkov11,Leibig93,Grebenkov13b}.

The computation of the Laplace-transformed heat kernel in
Sec. \ref{sec:kernel} is not limited to the considered basic domains
and can be extended to multi-layered structures such as a sequence of
parallel sheets, multiple co-axial cylinders of increasing radii, and
multiple concentric spherical shells.  For these structures, the
transport properties are supposed to be constant in each layer, while
transmission boundary conditions describe the exchange between
neighboring layers.  The radial eigenfunctions of the diffusion
operator are still expressed in terms of basic solutions $\I(z)$ and
$\K(z)$ in each layer, with the coefficients depending on the layer
(see \cite{Grebenkov10,Moutal19} for details).  As a consequence, one
can generalize the results from Sec. \ref{sec:kernel} to investigate
spectral sums in these domains that are often appear in modeling heat
transfer in composite materials, drug release from multilayered
capsules and drug-eluting stents, radioactive contamination of soils
and waste disposal,
etc. \cite{Siegel86,Shackelford91,Yuen94,Yates00,Pontrelli07,Hickson09,Liu09,Shackelford13}.
One can also consider diffusion in wedges, angular and spherical
sectors \cite{Redner,Thambynayagam}, as well as in the presence of
confining potentials \cite{Grebenkov15}.

In summary, we presented an overview of several mathematical tools for
computing explicitly the spectral sums involving zeros of linear
combinations of Bessel functions.  While each tool is fairly classical
and most of the derived summation formulas are known, these results
are dispersed in the literature.  As a consequence, as it is often
quite difficult to find out the needed tool or specific expression,
most authors re-derive these formulas from a scratch.  The recent book
by Thambynayagam \cite{Thambynayagam} partly resolved this difficulty,
providing on more than 2000 pages an exhaustive list of solutions of
the diffusion equation in all kinds of one- and two-dimensional
configurations.  In spite of its advantages, the book does not focus
on spectral sums, does not present the methodological part, and does
not discuss three-dimensional configurations.  In this light, the
present overview provides both a concise guide over rather generic
spectral sums and a practical recipe for evaluating the particular
ones.  These techniques are particularly valuable for getting explicit
solutions in some basic diffusion models and for investigating the
asymptotic behavior of the associated heat kernels.


\end{document}